\documentclass[onecolumn]{autart}    

\usepackage{graphicx}
\usepackage{textcomp}
\usepackage{lipsum}
\usepackage[english]{babel}
\usepackage{color,soul}
\usepackage{amsmath,amssymb,amsfonts}
\usepackage{subfig}
\newcommand*{\Scale}[2][4]{\scalebox{#1}{$#2$}}%
\usepackage{natbib}
\usepackage{algorithm}
\usepackage{url}

\usepackage{algorithmic}

\newcommand\smallmat[1]{\left[ \begin{smallmatrix}#1\end{smallmatrix}\right]}
\newcommand\bigmat[1]{\begin{bmatrix}#1\end{bmatrix}}

\newtheorem{thm_new}{Theorem}
\newtheorem{prop_new}{Proposition}
\newtheorem{cor_new}{Corollary}
\newtheorem{assum_new}{Assumption}
\newtheorem{rmk_new}{Remark}
\newtheorem{lemma_new}{Lemma}

\newcommand{\eq}[1]{\mbox{$ #1 $}}
\newcommand{\T}{^\top}

\usepackage{xcolor}

\newcommand{\revOne}[1]{{{\color{black} #1}}}
\newcommand{\revFive}[1]{{{\color{black} #1}}}
\newcommand{\revSeven}[1]{{{\color{black} #1}}}
\newcommand{\revCommon}[1]{{{\color{black} #1}}}

\newcommand{\RED}[1]{{{\color{black} #1}}}

\newif\ifred
\redfalse 

\begin{document}

\begin{frontmatter}
\title{Regional stability conditions for recurrent neural network-based control systems} 
\thanks[footnoteinfo]{Corresponding author.}
\author[Polimi]{Alessio La Bella\thanksref{footnoteinfo}}\ead{alessio.labella@polimi.it},    
\author[Polimi]{Marcello Farina}\ead{marcello.farina@polimi.it},               
\author[Polimi]{William D'Amico}\ead{william.damico@polimi.it},               
\author[Zack1,Zack2]{Luca Zaccarian}\ead{luca.zaccarian@laas.fr}               

\address[Polimi]{Dipartimento di Elettronica, Informazione e Bioingegneria, Politecnico di Milano, 20133, Milano, Italy.}                                  
\address[Zack1]{CNRS, LAAS, Universite de Toulouse, 31400 Toulouse, France}             
\address[Zack2]{Dipartimento di Ingegneria Industriale, University of Trento, 38122 Trento, Italy}             

\begin{abstract}                          
In this paper we propose novel global and regional stability analysis conditions based on linear matrix inequalities for a general class of recurrent neural networks. These conditions can be also used for state-feedback control design and a suitable optimization problem enforcing $\mathcal{H}_2$ norm minimization properties is defined. The theoretical results are corroborated by numerical simulations, showing the advantages and limitations of the methods presented herein.\\[-0.3cm]
\end{abstract}
\end{frontmatter}
\section{Introduction}\label{sec:intro}
Considering the growing availability of data in many engineering fields, there has been an increasing interest on neural networks~(NNs) and their properties in the last decades, in particular considering recurrent NNs~(RNNs). This is also true in the systems and control realm, where the latter have been subject of intense research in view of their useful features~\citep{sontag1992neural,lanzetti2019recurrent}. In particular, RNNs can be reliably used for modelling complex and nonlinear dynamical systems, given their stateful
nature that allows them to store memory of past data, as well as for data-based control design~\citep{bonassi2022recurrent}. The so-called \emph{indirect design methods}, for instance, consist of the design of suitable controllers
based on models learned from data and, in this context, RNN models are effective as they provide reliable approximations of nonlinear system dynamics. This approach has been adopted for the design of model predictive controllers, e.g., in \citep{armenio2019model,lanzetti2019recurrent}.
In this context, it is relevant to study systems-theoretic properties of RNNs, e.g., observability, controllability, and stability of motion. These properties are fundamental, e.g., for characterizing RNN-based models and the ensuing controller designs.
Different analysis conditions guaranteeing stability-related properties have been proposed. They include {global asymptotic stability} (GAS), {input-to-state stability}~(ISS), and {incremental} ISS ($\delta$ISS) \citep{jiang2001input,bayer2013discrete}. 
\ifred
Sufficient conditions ensuring GAS are derived for RNNs in~\citep{stipanovic2021stability}. Alternative contraction properties are characterized in \citep{buehner2006tighter} for echo state networks (ESNs) and in \citep{revay2020convex} for more general RNNs. Sufficient analysis conditions guaranteeing $\delta$ISS are studied in \citep{10178405, bonassi2021stabilitygru} for long short-term memory networks (LSTMs) and gated recurrent units (GRUs), respectively. Note that $\delta$ISS is a stronger stability property, as it implies ISS and GAS.  Thus, in \citep{WilliamArxiv} sufficient conditions ensuring  $\delta$ISS are proposed for a class of discrete-time nonlinear systems, which includes different common RNN classes, e.g., ESNs and neural nonlinear autoregressive networks with exogenous variables (NNARX).
\else
Sufficient conditions ensuring GAS are derived for RNNs in~\citep{stipanovic2021stability,hu2002global}, considering constant inputs. Alternative contraction properties are characterized in \citep{buehner2006tighter} for echo state networks (ESNs) and in \citep{revay2020convex} for more general RNNs. 
Sufficient analysis conditions guaranteeing ISS and/or $\delta$ISS are studied in \citep{armenio2019model, 10178405, bonassi2021stabilitygru,bonassi2021stability} for ESNs, long short-term memory networks (LSTMs), gated recurrent units (GRUs), and neural nonlinear autoregressive networks with exogenous variables (NNARX), respectively. Finally, in \citep{WilliamArxiv} sufficient conditions ensuring  $\delta$ISS are proposed for a class of discrete-time nonlinear systems, which includes  different common RNN classes, e.g., ESNs and NNARX. The conditions proposed in \citep{WilliamArxiv} appear to be less conservative and more suited for control design than previously-proposed ones, e.g., in \citep{bonassi2021stability,armenio2019model,hu2002global}.
\fi
\revSeven{Design conditions for NN-based controllers have been proposed in some relevant recent works~\citep{barabanov2002stability,liu2021overview}}. 
Design conditions for FFNN controllers are also provided in~\citep{vance2008discrete}, considering specific classes of second-order nonlinear systems.  More recently, event-triggered FFNN controllers have been proposed  in~\citep{de2023event_letters} for linear systems and  in~\citep{de2023event_auto} for perturbed ones, providing conditions to ensure local stability properties for the closed-loop system. \revOne{In this context, \citep{yin2021stability} developed conditions for stability analysis of closed-loop systems composed of a pre-designed FFNN-based controller and perturbed linear plant.}
The main limitations of results listed above, except for \citep{de2023event_letters}, \citep{de2023event_auto}, and \revOne{\citep{yin2021stability}}, is that they provide global stability properties. However, global stability conditions could be not always admissible. This may be due, e.g., to the RNNs nonlinear dynamics or to the boundedness of some state and input variables in a closed-loop scenario where the plant exhibits unstable dynamics. Similarly to what is done in~\citep{massimetti2009linear} and in~\citep{da2001local} in the case of linear plants with input saturation, it is necessary to develop regional stability conditions, associated with an estimate of the basin of attraction of the considered equilibrium.  \\
In view of the above discussion, this paper proposes novel global and local stability analysis conditions based on linear matrix inequalities (LMIs) for a general class of RNNs. These properties can be established thanks to the peculiarities of sigmoidal nonlinearities typically used in the RNN models. We also show that these conditions can be enforced in state-feedback control design. The generalization to more complex control systems (including, e.g., state observers, integrators, etc.) is rather straightforward and can be done following the lead of the present paper. The theoretical results are corroborated by numerical simulations, showing the advantages and limitations of the methods presented herein.\\
\ifred
The problem and the RNN model are introduced in Section~\ref{sec:ProbStat}. In Section~\ref{sec:RegCond}, a global stability condition for RNN models is derived and then, given its limitation, two different regional stability conditions are introduced. The derived conditions are then exploited for control design in Section~\ref{sec:controldesign}. Simulation results are illustrated in Section~\ref{sec:num} and conclusions are drawn in Section~\ref{sec:conclusion}.
\else
The problem is formulated in Section~\ref{sec:ProbStat}, where the considered RNN model is discussed and global stability conditions are provided. In Section~\ref{sec:RegCond}, two different regional stability conditions, based on different ideas, are introduced. The derived conditions are then used for control design in Section~\ref{sec:controldesign}. Simulation results are illustrated in Section~\ref{sec:num} and conclusions are drawn in Section~\ref{sec:conclusion}.%
\fi
%
\\\\\textit{Notation:} Let $\mathbb{R}$ denote the set of real numbers, $\mathbb{R}_{\geq 0}$ the set
 of positive or null real numbers including zero, and $\mathbb{R}_{> 0}$ the
 set of strictly positive real numbers. Given a matrix $A\in \mathbb{R}^{n\times n}$, its transpose is $A^\top$,  whereas its eigenvalues are $\lambda_i(A)$, with $i=$\revSeven{$1$}$\,\hdots,n$.  The entry in the $i$-th row and $j$-th column of a matrix $A$ is denoted $a_{ij}$. The $i$-th entry of a vector $v$ is indicated as $v_i$.  Given a matrix $P$, we use $P\succeq0$, \mbox{$P\succ0$}, \mbox{$P\preceq0$}, and \mbox{$P\prec0$} to indicate that it is positive semidefinite, positive definite, negative semidefinite, and negative definite, respectively. $\lambda_{\rm min}(P)$ and $\lambda_{\rm max}(P)$ denote the minimum and maximum eigenvalues of a symmetric matrix $P$, respectively. $0_{n,m}$ denotes a zero matrix with $n$ rows and $m$ columns and $I_n$ is the identity matrix of dimension $n$. Given a vector $v$, $\|v\|=\sqrt{v^Tv}$ denotes the 2-norm of a column vector $v$ and $\|v\|_Q=\sqrt{v^TQv}$ denotes the weighted Euclidean norm of $v$, where $Q$ is a positive definite matrix. Given a vector $v\in \mathbb{R}^n$, we define its infinity norm as $\|v\|_{\infty}=\max_{i=1,\cdots,n}(\|v_i\|)$. A matrix $A\in \mathbb{R}^{n\times n}$ is said to be Schur if $\|\lambda_i(A)\|<1$, for each $i=1,\cdots,n$. The set of positive definite symmetric real matrices is denoted as $\mathbb{S}^{n}_{\succ 0}= \{A \in \mathbb{R}^{n\times n}: A= A\T \wedge A \succ 0\}$, whereas the set of diagonal positive definite real matrices is denoted as $\mathbb{D}^{n}_{\succ 0}= \{A\in\mathbb{S}^{n}_{\succ 0}:\, a_{ij}= 0 \,\; \forall \,i,j\in[1,\hdots,n]\wedge i\neq j\}$.
\section{Problem statement}
\label{sec:ProbStat}
\ifred
\else
\subsection{{Plant model}}
\fi
\revSeven{This work is concerned with stability analysis and control design for a rather general class of Recurrent Neural Networks (RNNs) representing the plant model. We assume that the following discrete-time model, derived from a batch of input-output data $(u,z_{\rm m})$, accurately represents the plant dynamics.
\begin{subequations}
	\label{eq:sys1}
\begin{align}
	x^+&=A_{\circ}x+B_{u} u+B_{\sigma}\,\sigma(C_{\circ} x+D_u u)\label{eq:sys1_state}\\
	z_{\rm m}&=C_{\rm m}x\,,\label{eq:sys1_out}
\end{align}
\end{subequations}
\eq{x\in \mathbb{R}^n} is the state vector, \eq{u \in \mathbb{R}^m} the input vector, \eq{z_{\rm m} \in \mathbb{R}^p} the measurements vector}, $y\mapsto\sigma(y)=[\sigma_{1}(y_1)\cdots\sigma_{\nu}(y_\nu)]\T$ a vector of decentralized functions \eq{\sigma_i:\mathbb{R}\to\mathbb{R}} better specified below, \eq{A_{\circ}\in\mathbb{R}^{n\times n}}, \eq{B_u \in\mathbb{R}^{n\times m}}, \eq{B_{\sigma} \in \mathbb{R}^{n\times \nu}}, 
\eq{C_{\circ} \in\mathbb{R}^{\nu \times n}}, and
\eq{D_u\in\mathbb{R}^{\nu \times m}}. The following assumption is stated.

\begin{assum_new}\label{ass:1}
	Each component \eq{\sigma_i:\mathbb{R}\to\mathbb{R}} of function   $\sigma:\mathbb{R}^\nu\to\mathbb{R}^\nu$ in \eqref{eq:sys1}, i.e., with \eq{i\in \{1,\hdots,\nu\}}, is a sigmoid function, {namely, $\sigma_i$ is a monotone increasing, globally Lipschitz continuous with unitary Lipschitz constant, and such that $\sigma_i(0)=0$, $\sigma_i'(0)=\frac{d{\sigma_i(y_i)}}{dy_i} \Big \lvert_{y_i=0} =1$, and $\sigma_i(y_i) \in [-1,1]$ for all $y_i \in \mathbb{R}$.}
\end{assum_new}

\begin{figure}[b!]
	\centering
	\subfloat[\empty]{\includegraphics[trim=0 23 0 60,clip=true, width=0.35\linewidth]{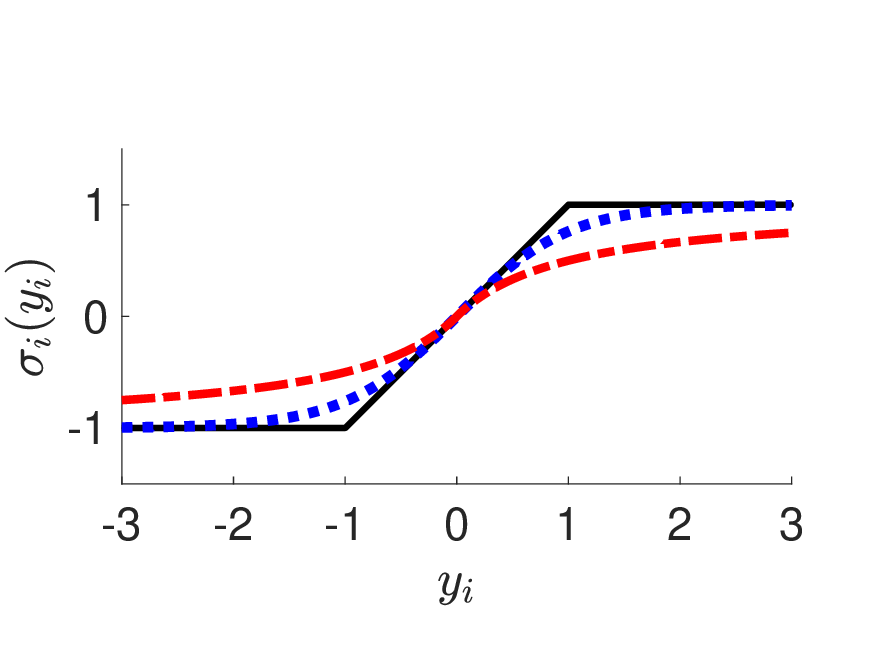}}
	\subfloat[\empty]{\includegraphics[trim=0 20 0 70,clip=true, width=0.35\linewidth]{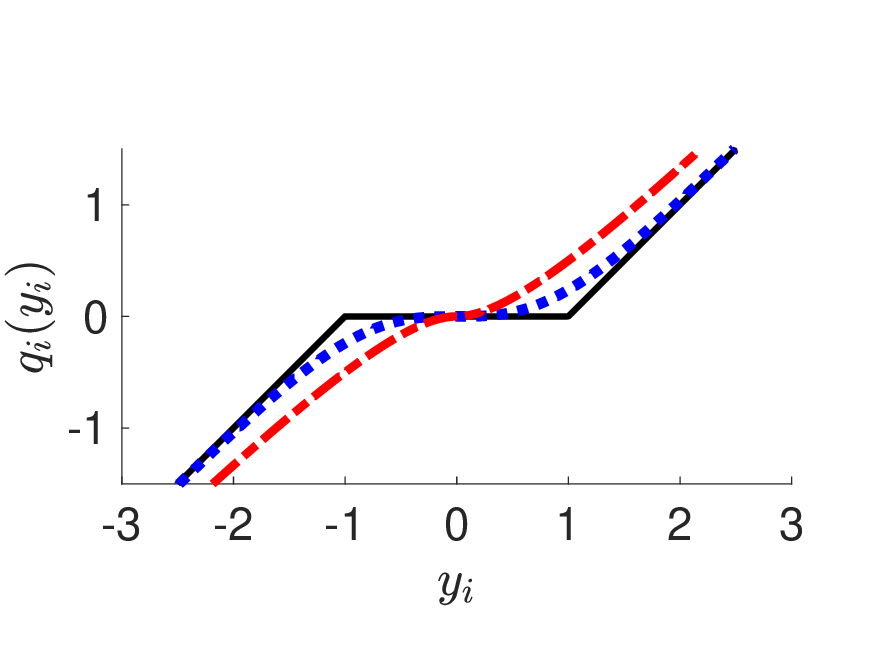}}
	\caption{Left plot: Examples of functions $\sigma_i$ satisfying Assumption 1: $\sigma_i(y_i)=\text{sat}(y_i)$ (solid black line), $\sigma_i(y_i)=\text{tanh}(y_i)$ (dotted blue line) and $\sigma_i(y_i)=y_i/(1+|y_i|)$ (dashed red line); Right plot: corresponding functions $q_i(y_i)$. }
	\label{fig:sigmoid}
\end{figure}

Assumption \ref{ass:1} is satisfied by different sigmoid functions, see Figure \ref{fig:sigmoid}.
For instance, model \eqref{eq:sys1} under Assumption~\ref{ass:1} represents generic classes of RNNs, e.g., Echo State Networks (ESNs) and shallow Neural Nonlinear AutoRegressive eXogenous (NNARX) networks. \revFive{The reader is referred to \citep[Section II]{WilliamArxiv} for more details on how the mentioned RNN families can be formulated as in \eqref{eq:sys1}}. \revFive{Note that  $\sigma$ can comprise different scalar components $\sigma_i$, provided that they all respect Assumption \ref{ass:1}.}
\smallskip\\
The purpose of this work is twofold. One goal is to establish LMI-based analysis conditions guaranteeing asymptotic stability of the origin for \eqref{eq:sys1} under linear state feedback $u=Kx$. Both global and local stability conditions will be proposed, with a possibly maximized ellipsoidal estimate of the basin of attraction. Then, the analysis conditions will be exploited for the design of a state feedback gain $K$, ensuring closed-loop stability and suitable performance. \revSeven{Note that the control law has state-feedback form just for the sake of simplicity, but a suitable observer should be used to provide state estimates through input-output data (this aspect will be later discussed in Remark 3); in fact, as usual in case of RNNs, the states of model \eqref{eq:sys1} may not correspond to real plant physical and measurable variables. The} resulting closed-loop system is
\begin{align}
	\label{eq:sys2}
	\begin{cases}
		\begin{split}
			\!\!\!\!	x^+&=Ax+B\,(y - \sigma(y))\\&= Ax+B\,q(y)\,,\\[0.1cm]
			y &= C x\,,
		\end{split}
	\end{cases}
\end{align}
where \eq{C=C_{\circ} +D_u K}, \eq{A = A_{\circ} + B_u K + B_{\sigma} C}, \eq{B = -B_\sigma}, and \eq{q:\mathbb{R}^\nu\mapsto \mathbb{R}^\nu} is a vector of scalar functions \linebreak \eq{y\mapsto q(y)=[q_1(y_1) \cdots q_\nu(y_\nu)]^\top} with \eq{q_i(y_i)=y_i -\sigma_i(y_i)}, for any $y_i \in \mathbb{R}$ and for each $i\in \{1,\hdots,\nu\}$. Figure \ref{fig:sigmoid} depicts \eq{\sigma_i(\cdot)} and \eq{q_i(\cdot)} for different functions respecting Assumption \ref{ass:1}, comparing them with two notable functions: the saturation $y\mapsto \text{sat}(y)=[\text{sat}_1(y_1) \cdots \text{sat}_\nu(y_\nu) ]$ and the deadzone  $y\mapsto\text{dz}(y) =y-\text{sat}(y)$, respectively, whose components are defined as
\ifred
{\mbox{$\text{sat}_i(y_i)=\max(-1,\min(y_i,1))$} and $\text{dz}_i(y_i) = y_i-	\text{sat}_i(y_i)$.}
\else 
\begin{subequations}\label{eq:sat_dy}
\begin{align}
	\text{sat}_i(y_i) &= 
	\begin{cases}
		1&\text{if $y_i\geq 1$}\\
		-1&\text{if $y_i\leq -1$}\\
		y_i&\text{otherwise}\,,
	\end{cases} \\[0.2cm]	\text{dz}_i(y_i) &= 
	\begin{cases}
		y_i-1 & \text{if $y_i\geq 1$}\\
		y_i+1 & \text{if $y_i\leq -1$}\\
		0 &\text{otherwise}\,.\\
	\end{cases}
\end{align}
\end{subequations}
\fi
Note that, under Assumption \ref{ass:1}, since \eq{\sigma(0)=0}, the origin is an equilibrium for \eqref{eq:sys2}. Thus, analysis and design conditions are here studied for $K$ ensuring exponential stability of the origin for the closed-loop system in \eqref{eq:sys2}.

\begin{rmk_new}
	The following results can be extended to plant models \eqref{eq:sys1} regulated by dynamic and nonlinear controllers provided that the control law can be expressed as a linear function of the states of the overall control system, as done in \citep{WilliamArxiv}.
\end{rmk_new}

\ifred
\revCommon{
\section{Stability Analysis}\label{sec:RegCond}
\subsection{Global stability analysis via saturation methods}\label{sec:GAS}
Considering global exponential
stability of system \eqref{eq:sys1}, a wide literature is available for the specific case  \eq{\sigma(\cdot)=\text{sat}(\cdot)}, where the sector characterization of this non-linearity is exploited \citep{massimetti2009linear,da2001local}. We show here that these conditions can be applied for any $\sigma$ under Assumption \ref{ass:1}. 
\begin{lemma_new}\label{lm:0}
	Let function $\sigma:\mathbb{R}^\nu\to\mathbb{R}^\nu$ satisfy Assumption \ref{ass:1} and consider $q:\mathbb{R}^\nu\to\mathbb{R}^\nu$ defined as \eq{y\mapsto q(y)= y- \sigma(y)}. Then, for any  $ y\in \mathbb{R}^\nu$ and \eq{W\in \mathbb{D}_{\succ 0}^{\nu}}, it holds that
\begin{align}
	\label{eq:sec_bound}
	q(y)\T\, W\, \sigma(y) \geq 0\,.\\[-0.8cm]\nonumber
\end{align}
\end{lemma_new}
The proof of Lemma \ref{lm:0} is reported in \citep{PaperExt} and its validity is notable from Figure \ref{fig:sigmoid}. In view of this result, the following stability condition can be stated.
\begin{prop_new}\label{prop:1}
	Under Assumption~\ref{ass:1}, the origin is a globally exponentially stable equilibrium for system \eqref{eq:sys2} if there exist \eq{S\in \mathbb{S}_{\succ 0}^{n}} and \eq{U\in \mathbb{D}_{\succ 0}^{\nu}}, satisfying the LMI
	\begin{align}\label{eq:globalLMI}
		&\begin{bmatrix}
			S &-SC^\top &SA^\top\\
			-CS & 2U & UB^\top\\
			AS & BU & S
		\end{bmatrix}\succ 0\,.\\[-0.8cm]\nonumber
	\end{align}
\end{prop_new}
The proof of Proposition~\ref{prop:1} derives directly from \citep{massimetti2009linear}, exploiting the fact that sector condition \eqref{eq:sec_bound} holds not only for the saturation case but for any function $\sigma$ satisfying Assumption \ref{ass:1}. Nevertheless, the global stability condition in Proposition~\ref{prop:1} can be too conservative in some cases, as characterized in the next lemma. 
\begin{lemma_new}\label{lm:1}
	There exists a solution to \eqref{eq:globalLMI}   only if both $A$ and $A_{\circ}+B_uK$ are Schur with a common quadratic Lyapunov function $V(x)=x^\top S^{-1} x$.
\end{lemma_new}
The proof Lemma \ref{lm:1} is straightforward from a Schur complement. An extensive proof is reported in \citep{PaperExt}.
Note that $A_{\circ}+B_uK$ is not Schur with closed loop schemes including an integral action, therefore condition \eqref{eq:globalLMI} can never hold in those cases, as discussed in~\citep{WilliamArxiv}. 
The intrinsic limitation highlighted by Lemma \ref{lm:1}  motivates the need for regional stability conditions for system \eqref{eq:sys2}, associated with an estimate of the basin of attraction of the stable equilibrium.

A sufficient condition guaranteeing regional stability of system \eqref{eq:sys2} is established in \citep{massimetti2009linear} for the saturation case, i.e., if \eq{\sigma(\cdot)=\text{sat}(\cdot)} and, as a consequence, \eq{q(\cdot)=\text{dz}(\cdot)}. The idea is to guarantee a Lyapunov decrease in a polytopic region characterized by \eq{\|H  x\|_{\infty} \leq 1}, where $H\in\mathbb{R}^{\nu\times n}$ is a further degree of freedom, used to enlarge, as much as possible, the estimate of the basin of attraction estimate of the origin. The key result for proving such a regional property derives from the fact the following sector characterization holds
\begin{align}\label{eq:sec_bound2}
	\text{dz}(y)\T\, W\, (\text{sat}(y) + Hx) \geq 0\,,
\end{align}
\eq{\forall y \in \mathbb{R}^\nu} and \eq{\forall x \in \mathcal{L}(H)=\{x \in \mathbb{R}^{n}: \|H  x\|_{\infty} \leq 1\, \}}. 
Nevertheless, condition \eqref{eq:sec_bound2} holds specifically for the saturation case and not for any function $\sigma$ satisfying Assumption \ref{ass:1}, thus implying that the regional stability results in \citep{massimetti2009linear} cannot be extended to our case.
}
\else
\subsection{Global exponential stability analysis}
\label{sec:GAS}
In this section we consider the problem of analysing global exponential
 stability of the origin for system \eqref{eq:sys2}. A wide literature is available for the specific case where \eq{\sigma(\cdot)} in \eqref{eq:sys2} is a saturation, e.g., \citep{massimetti2009linear,da2001local}, where the sector characterization of this non-linearity is exploited to obtain global stability conditions. 
In fact, for any \eq{y\in \mathbb{R}^\nu}, when \eq{\sigma(y)=\text{sat}(y)}, so that \eq{q(y)=y-\sigma(y)=\text{dz}(y)}, the well known sector $[0,I]$ condition ensures that  
\begin{align}
	\label{eq:sec_bound}
	q(y)\T\, W\, \sigma(y) \geq 0\,, 
\end{align}
for any  $ y\in \mathbb{R}^\nu$ and \eq{W\in \mathbb{D}_{\succ 0}^{\nu}}. In particular, in \citep{massimetti2009linear}, the sector condition \eqref{eq:sec_bound} is combined with quadratic Lyapunov stability inequalities to obtain LMI conditions ensuring global exponential stability of the origin for system \eqref{eq:sys2} with \eq{\sigma(\cdot)=\text{sat}(\cdot)}. Note that condition \eqref{eq:sec_bound} holds, not only for the saturation-deadzone case, but also for any function satisfying Assumption~\ref{ass:1}, as evident from Figure \ref{fig:sigmoid} and stated in the following. 

\begin{lemma_new}\label{lm:0}
Let function $\sigma:\mathbb{R}^\nu\to\mathbb{R}^\nu$ satisfy Assumption \ref{ass:1} and consider $q:\mathbb{R}^\nu\to\mathbb{R}^\nu$ defined as \eq{y\mapsto q(y)= y- \sigma(y)}. Then, condition \eqref{eq:sec_bound} holds for any  $ y\in \mathbb{R}^\nu$ and \eq{W\in \mathbb{D}_{\succ 0}^{\nu}}.
\end{lemma_new}
\textbf{Proof of Lemma \ref{lm:0}.}
To prove the stated lemma, it is sufficient to show that \eq{q_i(y_i)\, \sigma_i(y_i) \geq 0} holds for each $i\in \{1,\dots,\nu\}$, as the decentralized vector version of \eqref{eq:sec_bound} follows straightforwardly for any \eq{W\in \mathbb{D}_{\succ 0}^{\nu}}.\\
Consider a generic $i\in \{1,\dots,\nu\}$. From Assumption~\ref{ass:1}, it is evident that $\sigma_i(y_i)$ is globally Lipschitz continuous and, consequently, it is continuous for any $y_i\in \mathbb{R}$. Thus, also $q_i(y_i)=y_i-\sigma_i(y_i)$ is continuous for all $y_i\in \mathbb{R}$. For $y_i=0$, since $\sigma_i(0)=0$, it follows that $q_i(0)=0$. For $y_i>0$, it holds that $\sigma_i(y_i)>0$ as $\sigma_i(y_i)$ is monotone, and it is increasing in $y_i=0$, as  $\frac{d{\sigma_i(y_i)}}{dy_i} \Big \lvert_{y_i=0} =1$. Moreover, in view of the unitary Lipschitz constant of $\sigma_i$, it holds that $y_i \geq \sigma_i(y_i)$ and, consequently, that $q_i(y_i)\geq0$, for any $y_i>0$. Performing parallel steps, it follows that $\sigma_i(y_i)<0$ and $q_i(y_i)\leq0$, for any $y_i<0$. 
\hfill$\square$\\\\
Thus, the global stability results derived in \citep{massimetti2009linear} can be extended to the general form of \eqref{eq:sys2}, given that the same sector bound condition \eqref{eq:sec_bound} is applied.

\begin{prop_new}\label{prop:1}
	Under Assumption~\ref{ass:1}, the origin is a globally exponentially stable equilibrium for system \eqref{eq:sys2} if there exist \eq{S\in \mathbb{S}_{\succ 0}^{n}} and \eq{U\in \mathbb{D}_{\succ 0}^{\nu}}, satisfying the LMI
	\begin{align}\label{eq:globalLMI}
		&\begin{bmatrix}
			S &-SC^\top &SA^\top\\
			-CS & 2U & UB^\top\\
			AS & BU & S
		\end{bmatrix}\succ 0\,.
	\end{align}
\end{prop_new}
The proof of Proposition ~\ref{prop:1} is not reported here as it can be easily derived from \citep{massimetti2009linear}, exploiting the sector condition \eqref{eq:sec_bound}, which holds for any function satisfying Assumption \ref{ass:1}, as stated in Lemma \ref{lm:0}. Nevertheless, the global stability condition in Proposition~\ref{prop:1} is often too conservative to be of practical use. In fact, when \eqref{eq:sys2} stems from \eqref{eq:sys1} with $A_{\circ}+B_uK$ not being Schur, then condition \eqref{eq:globalLMI} can never hold, as for closed loop schemes including an integral action, as discussed in~\citep{WilliamArxiv}. 
\begin{lemma_new}\label{lm:1}
There exists a solution to \eqref{eq:globalLMI}   only if both $A$ and $A_{\circ}+B_uK$ are Schur with a common quadratic Lyapunov function $V(x)=x^\top S^{-1} x$.
\end{lemma_new}
\textbf{Proof of Lemma \ref{lm:1}.}
First of all, note that a necessary condition for \eqref{eq:globalLMI} is 
\begin{align}\label{eq:nec_cond}
	\begin{bmatrix}
		S   &S A^\top \\
	    AS  &S	\end{bmatrix}\succ0\,,
\end{align}
which implies, via congruence transformation \citep{boyd1994linear}, that 
\begin{align}\label{eq:congr_trans}
	\begin{bmatrix}
		S^{-1}   &A^\top S^{-1} \\
		S^{-1} A  &S^{-1}	\end{bmatrix}\succ0\,.
\end{align}
The Schur complement of \eqref{eq:congr_trans} is $A^\top S^{-1} A - S^{-1} \prec 0$, which holds if, and only if, $A$ is Schur, having quadratic Lyapunov function  $V(x)=x^\top S^{-1} x$, with $x \in \mathbb{R}^n$. 
\\
Now, consider that $A=A_{\circ}+B_uK+B_{\sigma}C$ and $B=-B_{\sigma}$ in~\eqref{eq:globalLMI}. Then, by pre- and post-multiplying~\eqref{eq:globalLMI} by 
\begin{align}\begin{bmatrix}
	I_n & 0 & 0\\ 0 & I_\nu & 0 \\ 0 & B_{\sigma} & I_n
\end{bmatrix}
\end{align} and its transpose, respectively, it follows that~\eqref{eq:globalLMI} is equivalent to
\begin{align}
	\label{pfmatr}
	\begin{bmatrix}
		S & -SC^{\top} & S(A_\circ+B_uK)^\top \\ 
		-CS & 2U & UB_\sigma^\top \\
		(A_\circ+B_uK)S & B_\sigma U & S
	\end{bmatrix}\succ0\,.
\end{align}
Similarly to before, exploiting the congruence transformation and the Schur complement, it follows that \eqref{pfmatr} implies that $(A_\circ+B_uK)^\top S^{-1} (A_\circ+B_uK)-S^{-1}\prec0$, which holds, if and only if, $A_\circ+B_uK$ is Schur stable, with quadratic Lyapunov function  $V(x)=x^\top S^{-1} x$, where $x \in \mathbb{R}^n$. 
\hfill$\square$\\\\
The intrinsic limitation highlighted by Lemma \ref{lm:1}  motivates the need to establish local stability conditions for system \eqref{eq:sys2}, or, even better, regional stability conditions associated with an estimate of the basin of attraction of the considered equilibrium.
\section{Regional stability analysis}
\label{sec:RegCond}
\subsection{Regional conditions with standard saturation}
A sufficient condition guaranteeing the regional stability of system \eqref{eq:sys2} has been established in the literature for the case where \eq{\sigma(\cdot)=\text{sat}(\cdot)} so that \eq{q(\cdot)=\text{dz}(\cdot)}. The idea is to guarantee a Lyapunov decrease in a polytopic region characterized by \eq{\|H  x\|_{\infty} \leq 1}, where $H\in\mathbb{R}^{\nu\times n}$ is a further degree of freedom, used to enlarge, as much as possible, the basin of attraction estimate for the zero equilibrium point. The key result for proving regional property was proposed by~\citep{da2005antiwindup}, stating that for any $W$$\in \mathbb{D}\revSeven{^{\nu}_{\succ 0}}$ the following holds:
\begin{align}\label{eq:sec_bound2}
	\text{dz}(y)\T\, W\, (\text{sat}(y) + Hx) \geq 0\,,\quad  \in \mathcal{L}(H),
\end{align}
\eq{\forall y \in \mathbb{R}^\nu} and \eq{\forall x \in \mathcal{L}(H)=\{x \in \mathbb{R}^{n}: \|H  x\|_{\infty} \leq 1\, \}}. Condition \eqref{eq:sec_bound2} can be seen as a non-global version of \eqref{eq:sec_bound} and it is often referred to as ``generalized sector condition". The following result can be proved using \eqref{eq:sec_bound2}.
\begin{prop_new}\label{cor:1} 
\normalfont{\textbf{\citep{massimetti2009linear}}} \textit{	The origin is a locally exponentially stable equilibrium for system \eqref{eq:sys2} with \eq{\sigma(\cdot)=\text{sat}(\cdot)} if there exist \eq{S\in \mathbb{S}_{\succ 0}^{n}}, \eq{U\in \mathbb{D}_{\succ 0}^{\nu}}, and \eq{L\in \mathbb{R}^{\nu\times n}}, satisfying the following LMI}
	\begin{subequations}\label{eq:zack_cond}
		\begin{align}\label{eq:zack_cond1}
			&\begin{bmatrix}
				S &-L\T-SC\T &SA\T\;\\
				-L-CS & 2U & UB\T\\
				AS & BU & S
			\end{bmatrix}\succ 0\,,\\ \nonumber \\[-0.3cm]
			&\begin{bmatrix}\label{eq:zack_cond2}
				S &L_{i,:}\T\\
				L_{i,:} & 1
			\end{bmatrix}\succeq 0\,, \quad \forall i \in \{1,\hdots, \nu\}\,,
			\\[-0.5cm] \nonumber 
		\end{align}
	where\ $L_{i,:}$ denotes the $i$-th row of matrix $L$.
\textit{		Moreover, the ellipsoidal set  \eq{\,\mathcal{E}(S)=\{x:{x}\T S^{-1}x\leq 1\}\subseteq \mathcal{L}(H)}, {with $H=LS^{-1}$}, is a forward invariant set contained in the basin of attraction of the origin.}
	\end{subequations}
\end{prop_new}

The proof of Proposition~\ref{cor:1} can be immediately derived from \citep{massimetti2009linear}. Specifically, condition \eqref{eq:zack_cond1} is obtained by combining a Lyapunov decrease inequality and the sector condition \eqref{eq:sec_bound2}. On the other hand, condition \eqref{eq:zack_cond2} implies that the sublevel set $\mathcal{E}(S)$ of function ${x}\T S^{-1}x$ is contained in the polytope $\mathcal{L}(H)$ where the sector condition \eqref{eq:sec_bound2} holds.
Comparing \eqref{eq:zack_cond} with~\eqref{eq:globalLMI}, we note that the additional decision variable $L$ (related to $H$) allows for more degrees of freedom in the LMI feasibility problem. {As a result,  no condition must be imposed on $A_{\circ}+B_uK$ for the feasibility of \eqref{eq:zack_cond}, thus overcoming the limitations of Lemma \ref{lm:1}. 
The main limitation of Proposition~\ref{cor:1} and of the generalized sector condition \eqref{eq:sec_bound2} is that they only hold for \eq{\sigma(\cdot)=\text{sat}(\cdot)} and they do not generally hold for functions $\sigma$ satisfying Assumption~\ref{ass:1}. 
\fi
This motivates the need of establishing novel regional stability conditions for a generic system \eqref{eq:sys2} satisfying Assumption \ref{ass:1}, representative of many RNN families. Thus, two novel regional stability conditions are proposed in the next two sections. The first one is associated with a single LMI condition, whereas the second one requires solving an iterative LMI-based algorithm. {The two conditions exhibit different advantages and uses,} as illustrated in Section \ref{sec:num}.
\subsection{Regional stability analysis via auxiliary function}
\label{sec:condition1}
\begin{figure}[t!]
	\centering
	\subfloat[\empty]{\includegraphics[trim=0 20 0 65,clip=true, width=0.35\linewidth]{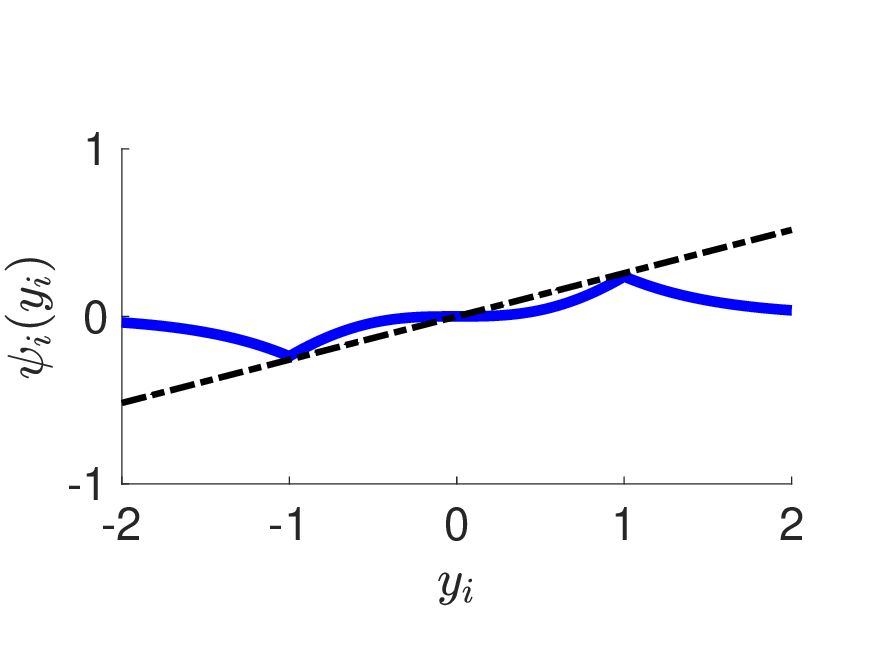}}
	\subfloat[\empty]{\includegraphics[trim=0 20 0 65,clip=true, width=0.35\linewidth]{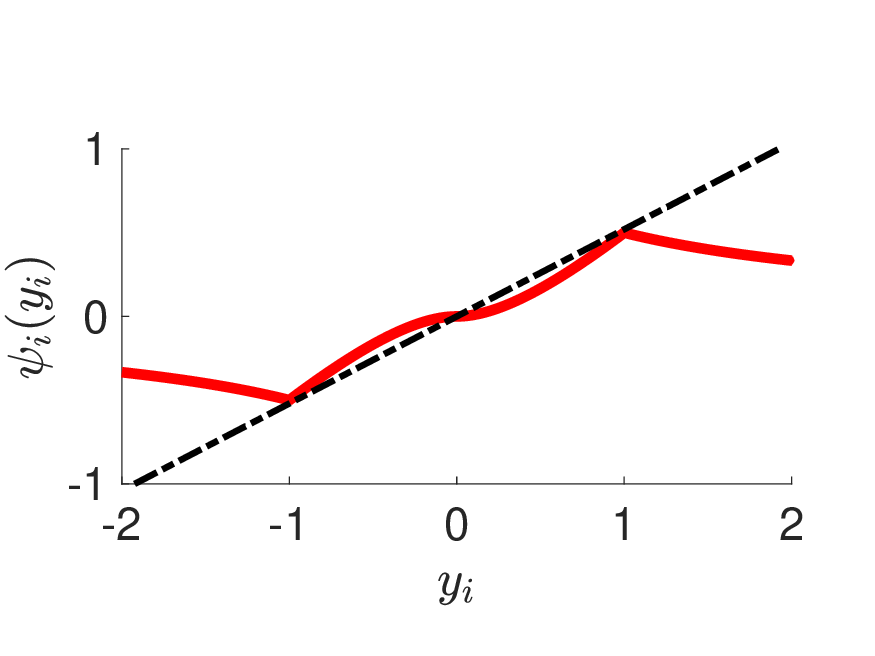}}
	\caption{{Function $\psi_i(y_i)$ (solid line) and the sector bound $\theta_i\,y_i$ (dashed black line). Left plot: $\sigma_i=\text{tanh}(y_i)$, with $\theta_i=0.2384$; Right plot: $\sigma_i(y_i)=y_i/(1+|y_i|)$, with $\theta_i=0.52$.\\} }
	\label{fig:psi}
\end{figure}

Consider system \eqref{eq:sys2}  under Assumption~\ref{ass:1} and 
define the auxiliary function 
\begin{align}\label{eq:psi}
	\psi(y) =  q(y) - \text{dz}(y)= \text{sat}(y)- \sigma(y).
\end{align}
Each element of the function $\psi$ corresponds to the difference between the saturation and the function $\sigma$. Introducing the function $\psi(y)$ has a twofold advantage. On one hand, it allows one to exploit the regional sector condition \eqref{eq:sec_bound2} by simply replacing $\text{dz}(y)$ by $q(y) -\psi(y)$ and by recalling that \eq{\text{sat}(y) =y-\text{dz}(y)}. On the other hand, the function \eq{\psi} allows establishing an additional sector condition, as evident from the next lemma.
\begin{lemma_new}\label{lm:3}
	Under Assumption \ref{ass:1}, function $y \mapsto \psi(y)$ in \eqref{eq:psi} belongs to the sector $[0,\Theta]$ for some \eq{\Theta \in \mathbb{D}_{\succ 0}^\nu} where $0 \preceq \Theta \prec I$. Equivalently, for any  $Y\in \mathbb{D}$\revSeven{$^\nu$}$_{\succ 0}$, it holds that 
	\begin{align}
		\label{eq:sec_bound5}
		\psi(y)\T\, Y\, (\Theta y-\psi(y))\geq 0\,, \quad \forall y \in \mathbb{R}^\nu.
	\end{align}
\end{lemma_new}
\ifred
The technical proof of Lemma \ref{lm:3} is reported in \citep{PaperExt}. The validity of condition \eqref{eq:sec_bound5}  is apparent from Figure \ref{fig:psi}, which reports the scalar component $\psi_i(y_i)$ for two different sigmoid functions $\sigma_i(y_i)$, together with the corresponding minimum sector bound $\theta_iy_i$.
\else
\textbf{Proof of Lemma \ref{lm:3}.}
First, being $\Theta$ and $Y$ diagonal, and being $y\mapsto\psi(y)=[\psi_{1}(y_1)\cdots\psi_{\nu}(y_\nu)]\T$ a vector of decentralized scalar functions, it is possible to focus on the scalar case, i.e., analysing a single $\psi_{i}:\mathbb{R}\to \mathbb{R}$. \\	Note that $\sigma_i(0)=0$ implies that $\psi_i(0)=0$. Due to the Lipschitz property of $\sigma_i$ and to the fact that \eq{\sigma_i(y_i) \in [-1,1]} for any $y_i$, it holds that
$|\sigma_i(y_i)|=|\sigma_i(y_i)-\sigma_i(0)|\leq\min\{1,|y_i|\}=|\text{sat}_i(y_i)|\,,$
	for any $ y_i \in \mathbb{R}$. Moreover, as discussed in the proof of Lemma \ref{lm:0}, the monotonicity of $\sigma_i$, the fact that $\sigma_i'(0)=1$, and $\sigma_i(0)=0$ imply that $y_i\sigma_i(y_i)\geq0$. As a consequence, $y_i\sigma_i(y_i)=|y_i||\sigma_i(y_i)|\leq|y_i||\text{sat}_i(y_i)|=y_i\text{sat}_i(y_i)$, thus implying that \eq{y_i\psi_i(y_i)=y_i(\text{sat}_i(y_i)- \sigma_i(y_i))\geq0,}  $\forall y_i \in \mathbb{R}$. 
Now,  introduce
\begin{align}\label{eq:theta2}
	\theta_i := \max_{y_i\in\mathbb{R}\setminus\{0\}} \frac{\psi_i(y_i)}{y_i}\,,
\end{align}
which represents the maximum slope of a line connecting the origin and $\psi_i(y_i)$, $\forall y_i \in \mathbb{R}\setminus\{0\}$. Note that \eq{\theta_i\geq0} because $y_i\psi_i(y_i)\geq0, \, \forall \,y_i \in \mathbb{R}$. On the other hand, it holds that $\frac{\sigma_i(y_i)}{y_i}>0, \, \forall \,y_i \in \mathbb{R}\setminus\{0\}$, because \eq{y_i\sigma_i(y_i)>0}, $ \forall y_i \in \mathbb{R}$, as discussed in the proof of Lemma \ref{lm:0}.
Thus, considering that $\frac{\text{sat}_i(y_i)}{y_i}\leq1, \, \forall \,y_i \in \mathbb{R}\setminus\{0\}$, the following holds
\begin{align*}
	\frac{\psi_i(y_i)}{y_i}={\frac{\text{sat}(y_i)}{y_i}}-\frac{\sigma_i(y_i)}{y_i}\leq1-{\frac{\sigma_i(y_i)}{y_i}}<1
\end{align*}
$\forall \,y_i \in \mathbb{R}\setminus\{0\}$, implying that $\theta_i<1$ in \eqref{eq:theta2}. Recalling that the function $\psi_i$ is bounded, then 
\begin{align}
\lim\limits_{y_i  \to +\infty}	\frac{\psi_i(y_i)}{y_i}=0\,,
\end{align}
and, by continuity, the ``max" in \eqref{eq:theta2} is attained at some finite value. Then,  it follows that $\frac{\psi_i(y_i)}{y_i}\leq\theta_i$, \mbox{$\forall\, y_i \in \mathbb{R}\setminus\{0\}$}, with $0\leq\theta_i<1$. As a result, it holds that $y_i(\theta_i y_i - \psi(y_i))\geq0$, $\forall\, y_i\in \mathbb{R}$. Moreover, since \eq{y_i\psi_i(y_i)\geq0}, \, $\forall\, y_i \in \mathbb{R}$, it holds that $\psi_i(y_i)(\theta_i y_i - \psi(y_i))\geq0$, $\forall\, y_i\in \mathbb{R}$. Finally, by defining $\Theta=\text{diag}(\theta_1,\dots,\theta_\nu)$, where $0\preceq\Theta\prec I$, it is apparent that $\psi(y)$ belongs to the sector bound $[0,\Theta]$ and that condition \eqref{eq:sec_bound5} holds for any $Y\in \mathbb{D}\revSeven{^{\nu}_{\succ 0}}$.
\hfill$\square$}

Note that {$\Theta=\text{diag}(\theta_1,\dots,\theta_\nu)$} is not a degree of freedom in \eqref{eq:sec_bound5}, but its value  is fixed and depends on the considered sigmoid function $\sigma(\cdot)$.  Each element $\theta_i$, $i=1,\dots,\nu$, can be computed numerically solving \eqref{eq:theta2}, e.g., by finding the maximum among the slopes of the lines connecting the origin to each point $(y_i,\psi_i(y_i))$. 
Figure \ref{fig:psi} reports the scalar component $\psi_i(y_i)$ for two different sigmoid functions $\sigma_i(y_i)$, providing the corresponding values of $\theta_i$ and illustrating the sector condition stated in Lemma \ref{lm:3}. 
\fi
Note that, for a given $\Theta$, the validity of Lemma \ref{lm:3} can be numerically certified through the COQ proof assistant \citep{chlipala2022certified}. Considering the functions depicted in Figure \ref{fig:psi}, the corresponding COQ proof code is available at \citep{COQfiles}. 
Before stating the main result, note that, in view of 
\revSeven{\eqref{eq:psi}}, the sector condition \eqref{eq:sec_bound2} can be rewritten as 
\begin{align}\label{eq:sec_bound3}
	(q(y) -\psi(y))\T W (y -  q(y) + \psi(y)+ Hx) \geq 0\,,
\end{align}
which, notably, holds for all \eq{x \in \mathcal{L}(H)} and all \eq{y \in \mathbb{R}^\nu}. \revFive{Note that the sector condition \eqref{eq:sec_bound2} only holds for the $\text{sat}(\cdot)$ nonlinearity and does not hold for a generic $\sigma$ satisfying Assumption \ref{ass:1}. Nevertheless, combining \eqref{eq:sec_bound2} with the sector condition \eqref{eq:sec_bound5}, which bounds the difference  between the saturation and the sigmoid function,  i.e., $\psi(\cdot)$, the novel regional stability condition \eqref{eq:sec_bound3} is enjoyed by system \eqref{eq:sys2} under Assumption \ref{ass:1}.  }

\begin{thm_new}\label{th:1}
	The origin is a locally exponentially stable equilibrium for system \eqref{eq:sys2} under Assumption \ref{ass:1} 
	if there exist \eq{S\in \mathbb{S}_{\succ 0}^{n}}, \eq{U\in \mathbb{D}_{\succ 0}^{\nu}}, \eq{R\in \mathbb{D}_{\succ 0}^{\nu}}, and \eq{L\in \mathbb{R}^{\nu\times n}}, satisfying the following linear matrix inequality system
	\begin{subequations}\label{eq:th1}
		\begin{align}\label{eq:th1_cond1}
			&\begin{bmatrix}
				S &-L\T-SC\T &-SC\T \Theta &SA\T\;\\
				-L-CS & 2U &0 & UB\T\\
				-\Theta C S  & 0 &2R & RB\T\\
				AS & BU & BR & S\\
			\end{bmatrix}\!\!\succ 0\,,\\ 
			&\begin{bmatrix}\label{eq:th1_cond2}
				S &L_{i,:}\T\\
				L_{i,:} & 1
			\end{bmatrix}\succeq 0, \quad \forall i \in \{1,\hdots, \nu\}\,, 
		\end{align}
	\end{subequations}
	where $\Theta$ is defined such that  \eqref{eq:sec_bound5} holds. Moreover, the set $\mathcal{E}(S)=$ $\{x:{x}\T S^{-1}x\leq 1\}$ satisfies $\mathcal{E}(S)\subseteq \mathcal{L}(H)$, with $H=LS^{-1}$, and it is a positively invariant set contained in the basin of attraction of the origin.
\end{thm_new}
\textbf{Proof of Theorem \ref{th:1}.}
\ifred
We first show that \eqref{eq:th1_cond2} implies $\mathcal{E}(S)\subseteq \mathcal{L}(H)$. Considering $P=S^{-1}$ with \eq{S\in \mathbb{S}_{\succ 0}^{n}}, \eqref{eq:th1_cond2} can be rewritten as
$
	\begin{bmatrix}
		S & 0 \\
		0& 1
	\end{bmatrix}
	\begin{bmatrix}
		P & H_{i,:}\T \\
		H_{i,:} & 1
	\end{bmatrix}\begin{bmatrix}
		S & 0 \\
		0& 1
	\end{bmatrix} \succeq 0\,,
$
which implies, by means of Schur complement, that
		\begin{align}\label{eq:bas_cond}
	{x}\T H_{i,:} \T H_{i,:} x   \leq	{x}\T Px \leq 1 , 
\end{align}
for any $x \in \mathcal{E}(S)$ and for all \eq{i \in \{1,\hdots, \nu\}}.
\else
We first show that \eqref{eq:th1_cond2} implies $\mathcal{E}(S)\subseteq \mathcal{L}(H)$. Considering $P=S^{-1}$ with \eq{S\in \mathbb{S}_{\succ 0}^{n}}, from \eqref{eq:th1_cond2}, it follows that
	\begin{align}\label{eq:lmi_sec}
		\begin{bmatrix}
			S &L_{i,:}\T\\
			L_{i,:} & 1
		\end{bmatrix}\,=	&\begin{bmatrix}
			S & 0 \\
			0& 1
		\end{bmatrix}
		\begin{bmatrix}
			P & H_{i,:}\T \\
			H_{i,:} & 1
		\end{bmatrix}\begin{bmatrix}
			S & 0 \\
			0& 1
		\end{bmatrix} \succeq 0\,,
	\end{align}
	and, equivalently, that
	\begin{align} \label{eq:Pconstr2}
		\begin{bmatrix}
			P & H_{i,:}\T \\
			H_{i,:} & 1
		\end{bmatrix}\succeq 0.
	\end{align}
	Thus, considering the Schur complement of \eqref{eq:Pconstr2},  for any $x \in \mathcal{E}(S)$ it follows that 
		\begin{align}\label{eq:bas_cond}
		{x}\T H_{i,:} \T H_{i,:} x   \leq	{x}\T Px \leq 1 , 
	\end{align}
	for all \eq{i \in \{1,\hdots, \nu\}}.
	\fi 
	Given the definition of $\mathcal{L}(H)$, then \eqref{eq:bas_cond} implies that $\mathcal{E}(S)\subseteq \mathcal{L}(H)$.
	Consider now the candidate Lyapunov function \eq{V(x) = x\T P x } for system \eqref{eq:sys2}. To prove the {local} stability of the origin in $\mathcal{E}(S)=\mathcal{E}(P^{-1})$, it is sufficient to show that $\Delta V(x)=V(x^+)-V(x)$, for any $x\in \mathcal{E}(S)\backslash \{0\}$, is
	\ifred
	\begin{align}
	\Delta V(x)\!= \!\widetilde{x}\T\!\! \left(\!\begin{bmatrix}
			A\T\\B\T\\0
		\end{bmatrix}\!P\!\begin{bmatrix}
			A&B&0
		\end{bmatrix}\!-\!\begin{bmatrix}
			P&0&0\\
			0&0&0\\
			0&0&0
		\end{bmatrix}\!\right)\!\widetilde{x}\!<\!0,\label{eq:local1_lyap}
	\end{align}
	\else
	\begin{align}
		&\Delta V(x)=V(x^+)-V(x)= \nonumber \\
		&=(Ax + Bq(Cx))\T P(Ax + Bq(Cx)) - x\T Px\nonumber =\\
		&= \widetilde{x}\T \left(\begin{bmatrix}
			A\T\\B\T\\0
		\end{bmatrix}\!P\!\begin{bmatrix}
			A&B&0
		\end{bmatrix}\!-\!\begin{bmatrix}
			P&0&0\\
			0&0&0\\
			0&0&0
		\end{bmatrix}\right)\widetilde{x}<0\,,\label{eq:local1_lyap}
	\end{align}
\fi
where \eq{\widetilde{x}\!=\![x\!\T \, q(Cx)\!\T \, \psi(Cx)\!\T ]\!\T}\!\!.	Given that $\mathcal{E}(S)\subseteq \mathcal{L}(H)$, the sector condition \eqref{eq:sec_bound3} can be exploited to prove the validity of \eqref{eq:local1_lyap}. To this end, \eqref{eq:sec_bound3} is rewritten as
	\begin{align}
		\label{eq:sector_bound3_2v}
		\widetilde{x}\T\begin{bmatrix}
			0&\Phi\T&-\Phi\T\,\\
			\Phi&-2W&2W\\
			-\Phi&2W&-2W
		\end{bmatrix}\widetilde{x}\geq 0\,,  \quad \forall\, x \in \mathcal{L}(H)\,,\end{align}
	where \eq{\Phi=W(C+H)} is introduced for the sake of compactness.
	Analogously, condition \eqref{eq:sec_bound5} can be rewritten as 
	\begin{align}
		\label{eq:sector_bound4_2v}
		\widetilde{x} \begin{bmatrix}
			0&0&\;C\T\!\Theta Y\,\\
			0&0&0\\
			Y\Theta C \;&0&-2Y
		\end{bmatrix}\widetilde{x} \geq 0\,.  \end{align}
	Thus, by leveraging the \textit{S}-procedure \citep{boyd1994linear}, it is possible to guarantee \eqref{eq:local1_lyap} under \eqref{eq:sector_bound3_2v} and \eqref{eq:sector_bound4_2v} if \eq{\exists \,\tau_1,\tau_2>0} such that
	\begin{align}
		&\begin{bmatrix}
			A\T\\B\T\\0
		\end{bmatrix}P\begin{bmatrix}
			A&B&0
		\end{bmatrix}-\begin{bmatrix}
			P&0&0\\
			0&0&0\\
			0&0&0
		\end{bmatrix}+\tau_1\!\begin{bmatrix}
		0&\Phi\T&-\Phi\T\,\\
		\Phi&-2W&2W\\
		-\Phi&2W&-2W
	\end{bmatrix}\nonumber\\[0.1cm] &+\tau_2\!\begin{bmatrix}
			0&0&C\T \Theta Y\,\\
			0&0&0\\
			Y\Theta C &0&-2Y
		\end{bmatrix}\!\!
		\prec0\,.\label{eq:s-proc_zack}
	\end{align}
	Now, introducing \eq{\tilde{W}=\tau_1 W}, \eq{\tilde{Y}=\tau_2 Y}, \revCommon{and $\tilde{\Phi}=\tau_1 \Phi = \tilde{W}(C+H)$}, by Schur complement, \eqref{eq:s-proc_zack} can be written as
	\begin{equation}
		\label{eq:DV2_02b}
		\begin{bmatrix}
			P&-\revCommon{\tilde{\Phi}}\T\;\;&\revCommon{\tilde{\Phi}}\T\!\!-C\T\Theta \tilde{Y}\;\;&A\T\\
			-\revCommon{\tilde{\Phi}}&2\tilde{W}&-2\tilde{W}&B\T \\
			\revCommon{\tilde{\Phi}}-\tilde{Y} \Theta C\;\;&-2\tilde{W}&2\tilde{W}+2\tilde{Y}&0\\
			A&B&0&P^{-1}
		\end{bmatrix}\succ0\,.
	\end{equation}
\ifred
Finally, consider \eq{U=\tilde{W}^{-1}}, \eq{R=\tilde{Y}^{-1}}, and the matrix
$	M_1= {	 \begin{bmatrix}
		S&0&0&0\\
		0&U&0&0\\
		0&R&R&0\\
		0&0&0&I_{n}
	\end{bmatrix}}.$
Pre and post-multiplying \eqref{eq:DV2_02b} by $M_1$ and its transpose, respectively, \eqref{eq:th1_cond1} is obtained.
\else
	Introducing \eq{U=\tilde{W}^{-1}}, \eq{R=\tilde{Y}^{-1}}, and the following auxiliary matrix
	\begin{align*}
		M_1=\begin{bmatrix}
			S&0&0&0\\
			0&U&0&0\\
			0&R&R&0\\
			0&0&0&I_{n}
		\end{bmatrix}\,,
	\end{align*}
	it follows that 
	\begin{align}
		M_1 \!&\begin{bmatrix}
			P&-\revCommon{\widetilde{\Phi}}\T\;\;&\revCommon{\widetilde{\Phi}}\T-C\T\Theta \tilde{Y}&A\T\\
			-\revCommon{\widetilde{\Phi}}&2\tilde{W}\;\;&-2\tilde{W}&B\T \\
			\revCommon{\widetilde{\Phi}}-\tilde{Y} \Theta C&-2\tilde{W}\;\;&2\tilde{W}+2\tilde{Y}&0\\
			A&B\;\;&0&P^{-1}
		\end{bmatrix} M_1\T=\nonumber \\[0.1cm]=&\begin{bmatrix}
			S &-L\T-SC\T &-SC\T \Theta &SA\T\;\\
			-L-CS & 2U &0 & UB\T\\
			-\Theta C S  & 0 &2R & RB\T\\
			AS & BU & BR & S\\
		\end{bmatrix}\!\!\succ 0\,,\label{eq:final_zack_cond}
	\end{align}
	where \eq{L=HS}. The validity of \eqref{eq:final_zack_cond} is guaranteed by \eqref{eq:th1_cond1}, being them equivalent. 
	\fi
	Thus, if \eqref{eq:th1} holds, \eqref{eq:local1_lyap} is verified for all $x\in \mathcal{E}(S)\backslash \{0\}$ and, consequently, the origin is locally exponentially stable with a basin of attraction including $ \mathcal{E}(S)$. 
\hfill$\square$
\\\\Comparing condition~\eqref{eq:th1_cond1} with~\eqref{eq:globalLMI}, it is worth noting that the additional decision variable $L$ (related to $H$) allows for more degrees of freedom in the LMI feasibility problem. As a result,  no condition must be imposed on $A_{\circ}+B_uK$ for the feasibility of \eqref{eq:th1_cond1}, thus overcoming the limitations of Lemma \ref{lm:1}.
\begin{lemma_new}\label{lm:feasth1}
	A necessary and sufficient condition for the 
		existence of a small enough $\Theta\succ0$
	ensuring feasibility of \eqref{eq:th1} is that $A$ be Schur stable.
\end{lemma_new}

\textbf{Proof of Lemma \ref{lm:feasth1}.}
The necessity that $A$ be Schur for \eqref{eq:th1_cond1} to be feasibile follows from the same argument as those in the proof of Lemma \ref{lm:1} and the inequality $\smallmat{S & SA^\top \\ AS & S} \succ 0$ following from \eqref{eq:th1_cond1}. 

Conversely, for the sufficiency, for any Schur $A$ there exists a positive definite $S$, which can be scaled to an arbitrarily small norm, such that $\smallmat{S & SA^\top \\ AS & S} \succ 0$. Now, given such a matrix $S$, select $L = -CS$ and note that \eqref{eq:th1_cond2} can be written as
\begin{align}
	\label{eq:proofLuca1}
	\bigmat{S & 0 \\ 0 & 1} \bigmat{S^{-1} & C_{i,:}^\top \\ C_{i,:} & 1} \bigmat{S & 0 \\ 0 & 1} \succeq 0, \quad 
	\forall i \in \{1, \ldots, \nu\},
\end{align}
which is clearly satisfied for a small enough scaling of $S$. Let us then assume, without loss of generality,
that the matrix $S$ satisfying $\smallmat{S & SA^\top \\ AS & S} \succ 0$ is sufficiently small to also satisfy 
\eqref{eq:proofLuca1}. 
\\
Consider now \eqref{eq:th1_cond1} and let us first consider the case with $\Theta = 0$. Due to the selection of $L=-CS$, then via a Schur complement, \eqref{eq:th1_cond1} holds if there exist $U \in {\mathbb D}^\nu_{\succ 0}$ and $R \in {\mathbb D}^\nu_{\succ 0}$ such that
\begin{align}
	\label{eq:withTheta0}
	&\bigmat{S & SA^\top \\ AS & S}\! -\! \bigmat{0 & 0 \\ BU & BR}\!\! \bigmat{(2U)^{-1} & 0 \\ 0 &(2R)^{-1}}\! \! \bigmat{0 & UB^\top \\ 0 & RB^\top} \!=	\nonumber \\
	& = \bigmat{S & SA^\top \\ AS & S} - \frac{1}{2}\bigmat{0 \\ B} (U+R) \bigmat{0 & B^\top} 
	\succ 0,
\end{align}
which clearly holds true for small enough selections of 
$U \in {\mathbb D}^\nu_{\succ 0}$ and $R \in {\mathbb D}^\nu_{\succ 0}$, due to the available margin in the (strict) positive definiteness of the matrix at the left.

Consider finally the case with $\Theta \succ 0$ and note that it corresponds to a continuous perturbation of the previous condition, which holds true, once again, due to the margin available from the positive definiteness 
of \eqref{eq:withTheta0},
and due to continuity of the eigenvalues as functions of the matrix entries.
\hfill$\square$

The condition provided in Theorem~\ref{th:1} is a powerful tool not only to guarantee local exponential stability of the origin for system \eqref{eq:sys2}, but also to maximize an estimate of the basin of attraction. This can be done, for example, by solving the following LMI optimization problem
	\begin{align}\label{eq:gamma_th1}
	\begin{split}
		&\underset{\substack{S\in \mathbb{S}_{\succ 0}^{n}, \,L\in \mathbb{R}^{\nu\times n},\\ U, R\, \in \mathbb{D}_{\succ 0}^{\nu}, \gamma\in \mathbb{R}} }{\max} \,\gamma\\
		&\text{subject to \eqref{eq:th1}\, and } f_{\rm LMI}(S)\succeq \gamma I\,,
	\end{split}
\end{align}
%
where $f_{\rm LMI}(S)$ can be selected by the designer based on the goal. For instance, as discussed in \cite[Section 5.2]{boyd1994linear}, if $f_{\rm LMI}(S)=\text{logdet}(S)$, the volume of the invariant ellipsoid \eq{\mathcal{E}(S)} will be maximized. On the other hand, if \eq{f_{\rm LMI}(S)=S}, the radius of the maximal ball inscribed in \eq{\mathcal{E}(S)} will be maximized. In both cases the optimization problem is convex. Alternative polytopic inner approximation can be also used. 
As a final remark, note that the use of two joint sector conditions, i.e., \eqref{eq:sec_bound2} and \eqref{eq:sec_bound3}, can make \eqref{eq:th1} \revOne{computationally heavy}, as it will be evident in Section \ref{sec:num}. For this reason, \revOne{a less complex}, but less computationally attractive, regional stability condition for system \eqref{eq:sys2} is proposed in the following section.
\subsection{Regional stability analysis via sector narrowing}
\label{sec:LAS1}
An alternative parametric sector condition enjoyed  by $\sigma(y)$ and $ q(y)=y-\sigma(y)$ can be provided, leading to the development of a different regional stability condition, described in the following lemma.
\begin{lemma_new}\label{lm:4}
	Under Assumption \ref{ass:1}, 
	there exists a monotone decreasing function $ \bar{y}_i:\mathbb{R}\to \mathbb{R}$, such that, for any $h_i>0$, it holds that 
		\begin{align}\label{eq:sec_polimi}
		y_i(\sigma_i(y_i)-h_iq_i(y_i))\geq0\,,\quad \forall \, y_i\in \mathcal{Y}_i(h_i)\,,
	\end{align}
	where $\mathcal{Y}_i(h_i)=[-\bar{y}_i(h_i),\;\bar{y}_i(h_i)]$.
	\end{lemma_new}
\begin{figure}[t!]
	\centering
	\includegraphics[trim=0 40 0 100,clip=true, width=0.35\linewidth]{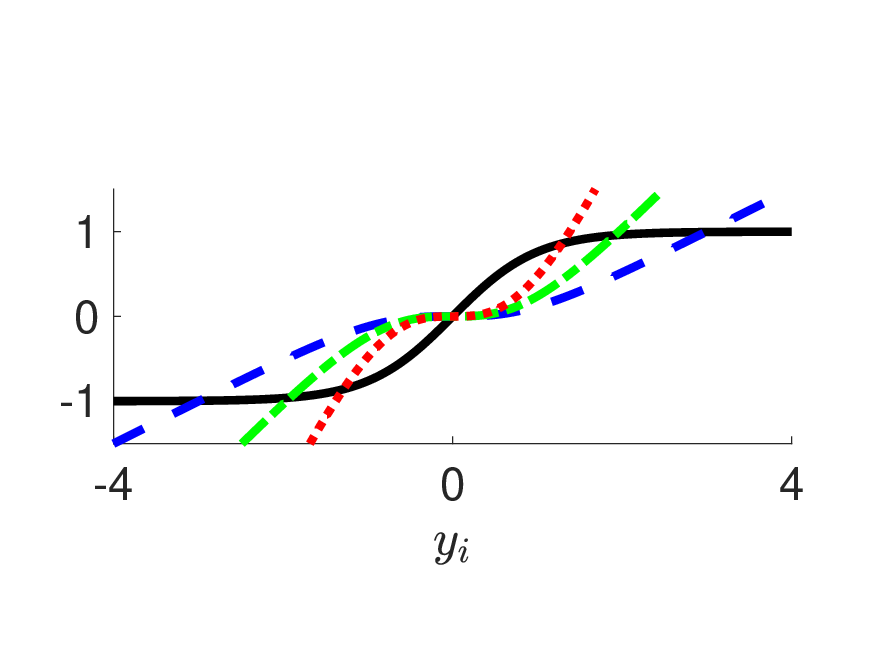}
	\caption{Function $\sigma_i(y_i)=\text{tanh}(y_i)$ (solid black) and the ensuing function $h_iq_i(y_i)$ for $h_i=0.5$ (dashed blue), $h_i=1$ (dashed-dotted green), $h_i=2$ (dotted red).}
	\label{fig:s-hq}
\end{figure}
\ifred
The validity of condition \eqref{eq:sec_polimi} is also illustrated in Figure \ref{fig:s-hq}, where the scalar function $\sigma_i(y_i)$ and the ensuing lower bound  $h_iq_i(y_i)$  are depicted for a specific sigmoid function and three specific values of $h_i$.
\else
	\textbf{Proof of Lemma \ref{lm:4}.} 
Since $q_i(y_i)=y_i-\sigma_i(y_i)$, then, for any $y_i\neq 0$, \eqref{eq:sec_polimi} can be rewritten as
	\begin{align}\label{eq:cond_h_i}
	y_i^2\,\Bigg(\frac{\sigma_i(y_i)}{y_i}(h_i+1) - h_i\Bigg)\geq0\,,
\end{align}
for any $ y_i \in \mathcal{Y}_i(h_i) \, \backslash\, \{0\}$. Introducing the function $\xi_i(y_i)=\sigma_i(y_i)/y_i$, condition \eqref{eq:cond_h_i} is equivalent to
\begin{align}\label{eq:cond_x_i}
	\xi_i(y_i)\geq \frac{h_i}{h_i+1}=\alpha_i(h_i)\,,\;\; \forall y_i \in \mathcal{Y}_i(h_i) \, \backslash\, \{0\}\,.
\end{align}
Given that both $\xi_i$ and $\alpha_i$ are continuous, $0<\alpha(h_i)<1$ for any $h_i>0$, and
\begin{subequations}\label{eq:lim_xi}
	\begin{align}
		&\;	\lim\limits_{y_i\to0^-}\,\xi_i(y_i)=1, \qquad \;	\lim\limits_{y_i\to0^+}\,\xi_i(y_i)=1,
	\\	&	\lim\limits_{y_i\to -\infty}\xi_i(y_i)=0,\qquad \lim\limits_{y_i\to +\infty}\xi_i(y_i)=0\,,\label{eq:lim12}
	\end{align}
\end{subequations}
then, for any $h_i>0$, there exists a non-empty \mbox{$\mathcal{Y}(h_i)=[-\bar{y}_i(h_i),\bar{y}_i(h_i)]$} satisfying \eqref{eq:cond_x_i}. The fact that  \eq{h_i\mapsto\bar{y}_i(h_i)} is monotone decreasing stems from the fact that $\frac{\text{d}\alpha_i(h_i)}{\text{d}h_i}>0$, and so condition \eqref{eq:cond_x_i} implies that, for each $\bar{h}_i > h_i$, it follows that $\alpha(\bar{h}_i) > \alpha(h_i)$ and, consequently the range of validity of \eqref{eq:cond_x_i} reduces, i.e., $\mathcal{Y}_i(\bar{h}_i)\subset \mathcal{Y}_i({h}_i)$.
	\hfill$\square$
\fi

Introducing  $\mathcal{Y}(H)=\mathcal{Y}_1(h_1)\times \cdots \times \mathcal{Y}_\nu(h_\nu)$, where $H=\text{diag}(h_1,\hdots,h_\nu)$,  \revFive{inequality \eqref{eq:sec_polimi} in} Lemma \ref{lm:4} \revFive{can be also expressed in matrix form. As a consequence}, 
for any $H \in \mathbb{D}_{\succ 0}^{\nu}$ and  any  $W \in \mathbb{D}_{\succ 0}^{\nu}$, there exists $\mathcal{Y}(H)$ such that
\begin{align}
\label{eq:sec_bound4}
q(y)\T W (\sigma(y) - H q(y)) \geq 0,\quad \forall \, y \in \mathcal{Y}(H).
\end{align}
In general, an explicit expression of $\mathcal{Y}(H)$ cannot be computed. However, a lower bound of $\mathcal{Y}(H)$ can be numerically computed based on the choice of $H$ and on the specific sigmoid function $\sigma$. \revCommon{To do that, for a given $H=\text{diag}(h_1,\hdots,h_\nu)$, it is sufficient to numerically compute the values of $y_i\neq0$  such that condition \eqref{eq:sec_polimi} holds with an equality, for each $i=1,\hdots,\nu$. Considering $\sigma_i$ to be an odd function for simplicity and given its sigmoidal structure defined in Assumption \ref{ass:1} (see Figure \ref{fig:s-hq}), two nonzero solutions exist, i.e.,  $y_i=\bar{y}_i(h_i)$ and $y_i=-\bar{y}_i(h_i)$, characterizing the bounds of  $\mathcal{Y}_i(h_i)$ for a given $h_i$.}
Also in this case, for given a $H$ and a corresponding set $\mathcal{Y}(H)$, the validity of \eqref{eq:sec_bound4} can be numerically certified exploiting the COQ assistant~\citep{chlipala2022certified}. 
\ifred
\else
Figure \ref{fig:s-hq} reports the scalar functions $\sigma_i(y_i)$ and the ensuing lower bound  $h_iq_i(y_i)$ for a specific sigmoid function and three specific values of $h_i$, well illustrating the sector condition \eqref{eq:sec_polimi} stated in Lemma \ref{lm:4}. 
\fi
The COQ code validating the sector condition for the considered $h_i$ in Figure \ref{fig:s-hq} is available at \citep{COQfiles}. Based on the regional sector condition \eqref{eq:sec_bound4}, the second main analysis result of this work can be stated.
%
	\begin{thm_new}\label{th:2}
		The origin is a locally exponentially stable equilibrium for system \eqref{eq:sys2} under Assumption \ref{ass:1}, 
		if there exists \eq{S\in \mathbb{S}_{\succ 0}^{n}}, \eq{U\in \mathbb{D}_{\succ 0}^{\nu}}, and $H=\text{diag}(h_1,\hdots,h_\nu)\in \mathbb{D}_{\succ 0}^{\nu}$,  such that
		\begin{subequations}\label{eq:th2}
			\begin{align}
				&\begin{bmatrix}\label{eq:th2_cond1}
					S &-SC\T &SA\T\\
					-CS & 2(H+I)U & UB\T\\
					AS & BU & S
				\end{bmatrix}\succ 0\\ 
				&\begin{bmatrix}\label{eq:th2_cond2}
					S &SC_{i,:}\T\\
					C_{i,:}S & \bar{y}^2_i(h_i)
				\end{bmatrix}\succeq 0, \quad \forall i \in \{1,\hdots, \nu\}\,, 
			\end{align}
		\end{subequations}
	with {$\bar{y}_i(h_i)$ defined such that \eqref{eq:sec_polimi} holds}. Moreover, the set \eq{\mathcal{E}(S)=\{x:{x}\T S^{-1}x\leq 1\}} is a forward invariant set and it is contained in the basin of attraction of the origin. Finally, for all $x\in \mathcal{E}(S)$, it holds that $y=Cx\in \mathcal{Y}(H)$.
	\end{thm_new}
\ifred
\textbf{Proof of Theorem \ref{th:2}}. Following similar steps to the ones before \eqref{eq:bas_cond}  and exploiting the definition of $\mathcal{E}(S)$, with \eq{P=S^{-1}\in \mathbb{S}_{\succ 0}^{n}},  \eqref{eq:th2_cond2}  implies that, for any \mbox{$x\in \mathcal{E}(S)$}, it holds that $\frac{{x}\T C_{i,:} \T C_{i,:} x } {\bar{y}^2_i(h_i)} \leq	{x}\T Px \leq 1$, 
for all \eq{i \in \{1,\hdots, \nu\}}. This shows that $y=Cx\in \mathcal{Y}(H)$ for all $x\in \mathcal{E}(S)$.
Consider now \eq{V(x) = x\T\, P\, x} as a candidate Lyapunov function. 
\else
\textbf{Proof of Theorem \ref{th:2}}. Following similar steps to \eqref{eq:lmi_sec}-\eqref{eq:bas_cond}  and exploiting the definition of $\mathcal{E}(S)$, with \eq{P=S^{-1}\in \mathbb{S}_{\succ 0}^{n}},  \eqref{eq:th2_cond2}  implies that, for any $x\in \mathcal{E}(S)$,
\begin{align}\label{eq:cond}
 \frac{{x}\T C_{i,:} \T C_{i,:} x } {\bar{y}^2_i(h_i)} \leq	{x}\T Px \leq 1,
\end{align}
	for all \eq{i \in \{1,\hdots, \nu\}}. This shows that $y=Cx\in \mathcal{Y}(H)$ for all $x\in \mathcal{E}(S)$.
	Consider now \eq{V(x) = x\T\, P\, x} as a candidate Lyapunov function. 
\fi
\ifred
Local exponential stability of the origin in $\mathcal{E}(S)=\mathcal{E}(P^{-1})$ holds if, $ \forall x \in \mathcal{E}(S)\,\backslash\,\{0\},$
		\begin{align}\label{eq:local_lyap_polimi}
			&\Delta V(x)=V(x^+) - V(x) =\nonumber\\
			&\begin{bmatrix}
				x\!\T \,q(Cx)\!\T\!
			\end{bmatrix}\!
			\left(\!\begin{bmatrix}\!
				A\!\T\\B\!\T
			\end{bmatrix}\!P\!\begin{bmatrix}
			\!	A&B
			\end{bmatrix}\!-\!\begin{bmatrix}
				\!P& 0\\
			\!	0&0
			\end{bmatrix}\!\right)\!\!
			\begin{bmatrix}
			\!	x\\q(Cx)\!
			\end{bmatrix}\!\!<\!0.
	\end{align}
	\else
	Local exponential stability of the origin in $\mathcal{E}(S)=\mathcal{E}(P^{-1})$ holds if 
	\begin{align}\label{eq:local_lyap_polimi}
			&\Delta V(x)=V(x^+) - V(x) =\nonumber\\&= (Ax + Bq(Cx))\T P(Ax + Bq(Cx)) - x\T Px\nonumber
			\\[0.1cm]&=\begin{bmatrix}
				x\T \,q(Cx)\T
			\end{bmatrix}\!
			\left(\begin{bmatrix}
				A\T\\B\T
			\end{bmatrix}P\begin{bmatrix}
				A&B
			\end{bmatrix}-\begin{bmatrix}
				P& 0\\
				0&0
			\end{bmatrix}\right)\!
			\begin{bmatrix}
				x\\q(Cx)
			\end{bmatrix}\nonumber\\&<0\,, \qquad \qquad  \forall x \in \mathcal{E}(S)\,\backslash\,\{0\}\,.
	\end{align}
\fi 
\ifred
Given $y=Cx\in \mathcal{Y}(H)$ for all $x\in \mathcal{E}(S)$, \eqref{eq:sec_bound4} can be exploited to prove \eqref{eq:local_lyap_polimi}, by imposing that
$
\Delta V(x)+q(Cx)\T\! W\! (Cx-	q(Cx) - H q(Cx)) \!<\!0,
$
for any $x\in \mathcal{E}(S)\,\backslash\{0\}$. By developing the quadratic form of the latter inequality, it follows that \eqref{eq:local_lyap_polimi} holds if
$\begin{bmatrix}
		A\T\\B\T
	\end{bmatrix}P\begin{bmatrix}
		A&B
	\end{bmatrix}-\begin{bmatrix}
		P& 0\\
		0&0
	\end{bmatrix}+\,\begin{bmatrix}
		0&C\T W\\WC&-2W(H+I)
	\end{bmatrix}\prec 0\,,
$
which, by a Schur complement, holds if, and only if, 
\begin{equation}\label{eq:DV01matrix}
	\begin{bmatrix}
		P&- C^\top{W}&A\T\\
		-{W}C&2{W}(H+I)&B\T\\
		A&B&P^{-1}
	\end{bmatrix}\succ0\,.
\end{equation}
\else
Given $y=Cx\in \mathcal{Y}(H)$ for all $x\in \mathcal{E}(S)$, \eqref{eq:sec_bound4} can be exploited to prove \eqref{eq:local_lyap_polimi}, by imposing that
	\begin{align}\label{eq:lyap_sec}
	\Delta V(x)+q(Cx)\T\! W\! (Cx-	q(Cx) - H q(Cx)) \!<\!0,
\end{align}
for any $x\in \mathcal{E}(S)\,\backslash\{0\}$. By developing the quadratic form of \eqref{eq:lyap_sec}, as in \eqref{eq:local_lyap_polimi}, it follows that \eqref{eq:local_lyap_polimi} holds if
\begin{equation}
	\label{eq:DV01}\begin{bmatrix}
		A\T\\B\T
	\end{bmatrix}P\begin{bmatrix}
		A&B
	\end{bmatrix}-\begin{bmatrix}
		P& 0\\
		0&0
	\end{bmatrix}+\,\begin{bmatrix}
		0&C\T W\\WC&-2W(H+I)
	\end{bmatrix}\prec 0\,.
\end{equation}
By a Schur complement, 	\eqref{eq:DV01} holds if, and only if, 
\begin{equation}\label{eq:DV01matrix}
	\begin{bmatrix}
		P&- C^\top{W}&A\T\\
		-{W}C&2{W}(H+I)&B\T\\
		A&B&P^{-1}
	\end{bmatrix}\succ0\,.
\end{equation}
\fi
\ifred
Finally, define  \eq{U={{W}}^{-1}} and \eq{M_2=\text{diag}(S,U,I)}. Then, pre and post-multiplying \eqref{eq:DV01matrix} by $M_2$ and its transpose, \eqref{eq:th2_cond1} is obtained.
\else
Defining  \eq{U={{W}}^{-1}} and \eq{M_2=\text{diag}(S,U,I)},
it follows that
\begin{align}
	\begin{split}\label{eq:DV02}
		&M_2\T
		\begin{bmatrix}
			P&- C\T{W}&A\T\\
			-{W}C&2{W}(H+I)&B\T\\
			A&B&P^{-1}
		\end{bmatrix}	
		M_2=\\&=
		\begin{bmatrix}
			S &-SC\T &SA\T\\
			-CS & 2(H+I)U & UB\T\\
			AS & BU & S
		\end{bmatrix}\succ 0\,.
	\end{split}
\end{align}
\fi
Therefore, \eqref{eq:th2} implies the validity of \eqref{eq:local_lyap_polimi}, proving local exponential stability of the origin with basin of attraction containing $\mathcal{E}(S)$. 
\hfill$\square$

Also in this case, the limitation of Lemma \ref{lm:1} are overcome by means of regional stability analysis. The following lemma provides a necessary and sufficient condition for the feasibility of the matrix inequality~\eqref{eq:th2}.
\begin{lemma_new}\label{prop:2}
A necessary and sufficient condition for the feasibility of \eqref{eq:th2} is that $A$ is Schur stable.
\end{lemma_new}
\ifred
\else
\textbf{Proof of Lemma \ref{prop:2}.} 
\ifred
\RED{Pre-multiplying and post-multiplying \eqref{eq:th2_cond1} by
	\begin{align}
		M_3 = 	\begin{bmatrix}
			I_n & 0 & 0\\
			0 & 0 & I_n\\
			0 & I_{\nu} & 0
		\end{bmatrix}\,,
\end{align}
and its transpose, respectively, the following obtained
	\begin{align}\label{eq:prop2_cond}	
		\begin{bmatrix}
			E_{11}&E_{12}\\
			E_{12}\T&E_{22}
		\end{bmatrix}\succ0\,
\end{align}
	where
$
	E_{11}=\begin{bmatrix}
		S&SA\T\\
		AS&S
	\end{bmatrix}$, $
	E_{12}=
	\begin{bmatrix}
		-SC\T\\
		BU\end{bmatrix}\!$, and \mbox{$E_{22}=2HU+2U$}.
}\else
Introducing
	\begin{align}
		M_3 = 	\begin{bmatrix}
			I_n & 0 & 0\\
			0 & 0 & I_n\\
			0 & I_{\nu} & 0
		\end{bmatrix}\,,
	\end{align}
	it is evident that \eqref{eq:th2_cond1} is equivalent to 
	\begin{align}
		\begin{split}
			\label{eq:prop2_cond}
			M_3
			\begin{bmatrix}
				S&-SC\T&SA\T\\
				-CS&2(HU+U)&UB\T\\
				AS&BU&S
			\end{bmatrix} 		
			M_3\T
			=	\begin{bmatrix}
				E_{11}&E_{12}\\
				E_{12}\T&E_{22}
			\end{bmatrix}\succ0\,
		\end{split}
	\end{align}
	where
\begin{align*}
	E_{11}=\begin{bmatrix}
		S&SA\T\\
		AS&S
	\end{bmatrix}\!, 
	E_{12}=
	\begin{bmatrix}
		-SC\T\\
		BU\end{bmatrix}\!, E_{22}=2HU+2U.
\end{align*}
\fi
	Recalling that \eq{P=S^{-1}} and in view of the Lyapunov stability inequality, a matrix \eq{S\succ0} exists satisfying \eq{E_{11}\succ0}, if, and only if, $A$ is Schur stable, which proves the necessity condition in Lemma \ref{prop:2}. 
	To prove sufficiency, select $S_\circ$ such that \eq{E_{11}\succ0} with $S=S_\circ$ (as well as with $S=\mu S_\circ$ for any $\mu >0$). Moreover, select any $U_\circ$ and a large enough $H\succ0$ such that \eqref{eq:prop2_cond} with $S=S_\circ$ and $U=U_\circ$ is satisfied. Such a large enough selection of $H$ exists because \eq{E_{11}\succ0} and \eq{E_{22}} can be made arbitrarily large by adjusting $H$. 
	Note now that, with the selected $H$, \eqref{eq:th2_cond1}, being a homogeneous matrix in $S$ and $U$, which holds for any selection $S=\mu S_\circ$ and $U=\mu U_\circ$ with $\mu>0$. Finally, to satisfy \eqref{eq:th2_cond2}, it is enough to choose $\mu>0$ sufficiently small in the selection of $S=\mu S_\circ$. This is always possible as, by following similar steps to \eqref{eq:lmi_sec}-\eqref{eq:bas_cond}, \eqref{eq:th2_cond2} becomes 
	\begin{align} \label{eq:Pconstr}
		\begin{bmatrix}
			S^{-1} & C_{i,:}\T \\
			C_{i,:} & \bar{y}_i^2(h_i)
		\end{bmatrix}\succeq 0\,,
	\end{align}
where $H\succ 0$ and, consequently, $\bar{y}_i^2(h_i)$ have been fixed. The selection of $(H,\mu S_\circ,\mu U_\circ)$  then satisfies \eqref{eq:th2}.
\hfill$\square$
\fi
Following Lemma \ref{prop:2}, the next corollary holds.
\begin{cor_new}\label{cor:2}
	If $A$ is Schur stable, a scalar $\bar{h}>0$ exists such that \eqref{eq:th2} admits a feasible solution for any \eq{H \succeq \bar{h}I}. 
\end{cor_new}
\ifred
\else
\textbf{Proof of Corollary \ref{cor:2}}. 
If $A$ is Schur stable, the existence of a large enough selection $\bar{H}$ of $H$ (and the ensuing selections $\bar{S}$ and $\bar{U}$ of $S$ and $U$, respectively) satisfying \eqref{eq:th2} is proven in Lemma \ref{prop:2}. Thus, it is possible to fix  $\bar{h}=\lambda_{\text{min}}(\bar{H})$. In fact, for any selection $H\succ \bar{h}I$, with $H\in \mathbb{D}^\nu_{\succ 0}$, the homogeneous constraint \eqref{eq:th2_cond1} is satisfied by any  $S=\bar{\mu}\bar{S}$ and  $U=\bar{\mu}\bar{U}$, with $\bar{\mu}>0$. Additionally, for any $H\succ \bar{H}$, with $H\succ \bar{h}I$, with $H\in \mathbb{D}^\nu_{\succ 0}$, considering the set $\mathcal{Y}(H)$, the validity of \eqref{eq:th2_cond2} can be proven using similar arguments to the proof of Lemma \ref{prop:2}, selecting $S=\bar{\mu}\bar{S}$ and a sufficiently small $\bar{\mu}>0$.
\hfill$\square$
\fi

Since $\bar{y}_i(\cdot)$ in Lemma \ref{lm:4} is a monotone decreasing function,
\revFive{it is expected that large estimates of the basin of attraction are achieved close to small values of $H$ (and of $\bar{h}$), since these  correspond to large regional sectors $\mathcal{Y}(H)$.}
The value $\bar{h}$ constructed in Corollary~\ref{cor:2} may  be conservative in terms of size of the estimated basin of attraction. It is therefore suggested to select an optimized $H$ according to the following LMI optimization problem
\begin{subequations}\label{eq:minh_th2}
		\begin{align}
			&	\underset{\substack {S\in \mathbb{S}_{\succ 0}^{n}, U\in \mathbb{D}_{\succ 0}^{\nu},\\ H_u\in\mathbb{D}_{\succ 0}^{\nu}, \gamma_u  \in \mathbb{R}}}{\min} -\gamma_u \label{eq:cf_minH}\\
			\text{subject}&\text{ to} \nonumber \\
			&\begin{bmatrix}\label{eq:cond_minH}
				S&-SC\T&SA\T\\
				-CS&2(H_u+U)&UB\T\\
				AS&BU&S
			\end{bmatrix} 	\succ 0,\\
			& \;\,U \succeq I\,,\label{eq:Umin}\\
			& \;H_u \preceq \gamma_u I\label{eq:gammaubound}\,.
		\end{align}
\end{subequations}
\!\! where $H_u=HU$ is introduced. Note that~\eqref{eq:Umin} is imposed (in place of $U \succ 0$) in view of the fact that \eqref{eq:cond_minH} is homogenous in the decision variables $(U,H_u,S)$. In fact, one may select $\bar{h}=\gamma_u$ in Corollary \ref{cor:2} because any feasible solution to \eqref{eq:minh_th2} satisfies $H=H_uU^{-1}\preceq H_u\preceq \gamma_u I$, so that any larger selection of $H$ is a feasible one. More generally, one may obtain a potentially less conservative estimate of the minimum $\bar{h}$ by selecting $\bar{h}=\lambda_{max}(H_u U^{-1})\leq \gamma_u$.\smallskip\\
%
As done in Section~\ref{sec:condition1}, it is possible to maximize the size of the basin of attraction estimate $\mathcal{E}(S)$ , introduced in Theorem \ref{th:2}. Given $H$ (possibly selected according to \eqref{eq:minh_th2}), similarly to \eqref{eq:gamma_th1}, the size of $\mathcal{E}(S)$ can be maximized by solving the following optimization problem
%
	\begin{align}\label{eq:gamma_th2}
		\begin{split}
			&\underset{\substack {S\in \mathbb{S}_{\succ 0}^{n}, \,U\in \mathbb{D}^{\nu}_{\succ 0},\\ \gamma \in \mathbb{R}}}{\max} \,\gamma\\
			&\text{subject to \eqref{eq:th2}\, and } f_{\rm LMI}(S) \succeq \gamma I\,.
		\end{split}
	\end{align}
for different possible functions $f_{\rm LMI}(S)$, as discussed in Section \ref{sec:condition1}.
Overall, the estimate $\mathcal{E}(S)$ can be maximized  through the procedure described in Algorithm~\ref{alg:1}, which iteratively adjusts the selection of $H$ \revFive{such that the largest estimate of the basin of attraction is computed}. 
\begin{algorithm}[t!]
	\caption{Optimization of $\mathcal{E}(S)$ through \eqref{eq:th2}}
	\begin{algorithmic}[1]\label{alg:1}\vspace*{0.1cm}
		\ENSURE $A$ in system \eqref{eq:sys2} is Schur stable \\[0.1cm]
		\STATE	\textbf{Set} $i=0$ and $\Delta h>0$ to a suitable small parameter \\[0.1cm]
		\STATE	\textbf{Solve}  \eqref{eq:minh_th2} and select $\bar{h}=\lambda_{max}(H_u U^{-1})$ \\[0.1cm]
		\WHILE{$i\leq i_{\text{max}}$}  \vspace*{0.1cm} 
		\STATE\textbf{Set} $H = (\bar{h} + i\Delta h)I$ \vspace*{0.1cm}
		\STATE \textbf{Solve} \eqref{eq:gamma_th2}, and denote the optimal solution as \eq{S^{(i)}}, $\gamma^{(i)}$  \\ \vspace*{0.1cm}
		\STATE  \textbf{Set}  $i=i+1$\vspace*{0.1cm} \\
		\ENDWHILE \vspace*{0.1cm} \\
		\STATE \textbf{Select} $i_{\text{best}}=\underset{i=0,\hdots,i_{\text{max}}}{\text{argmax}}\gamma^{(i)}$\\\vspace*{0.1cm}
		The best estimate of the basin of attraction of the origin is $\mathcal{E}(S^{(i_{\text{best}})})$.
	\end{algorithmic}
\end{algorithm}
\revFive{In particular, Algorithm~\ref{alg:1} is structured as follows. Initially, \eqref{eq:minh_th2} is solved to find the minimum value of $\bar{h}$ such that \eqref{eq:th2} is feasible. Then, $H$ is iteratively increased for a fixed number of iterations (i.e., $i_\text{max}$) and the estimate of the basin of attraction is computed at each iteration. Ultimately, the iteration corresponding to the largest estimate of the basin of attraction is selected.}
\subsection{Choice of the regional stability condition} \label{sec:choice}
The regional condition based on the auxiliary function (i.e.,~\eqref{eq:th1}) is recommended when wanting a \revOne{prompt} application. Indeed,~\eqref{eq:th1} is associated with a single LMI problem, \revOne{i.e., \eqref{eq:gamma_th1},} whereas when the sector narrowing conditions~\eqref{eq:th2} are exploited, more cumbersome iterative LMIs must be solved to find the matrix $H$ \revOne{returning the largest estimate of the basin of attraction} (cf. Algorithm~\ref{alg:1}), which may be time consuming. On the other hand, condition~\eqref{eq:th2} and Algorithm~\ref{alg:1} are preferred when the user aims to \revOne{deal with large-order systems}. Indeed, as shown in Section \ref{sec:num},  the condition based on the auxiliary function (i.e.,~\eqref{eq:th1}) leads to numerical problems in case of \revOne{high-dimensional systems}, while the one based on sector narrowing (i.e.,~\eqref{eq:th2}) is still feasible, corroborating the potential of the latter condition. \revOne{On the other hand, results in Section \ref{sec:num} show that, when both regional stability conditions are feasible, the one based on the auxiliary function, i.e., \eqref{eq:th1}, returns a larger estimate of the basin of attraction.  \\
Finally, it is worth noting that, for both conditions, the dimension of the considered regional sectors, i.e., $\mathcal{L}(H)$ for the LMI problem \eqref{eq:gamma_th1} and $\mathcal{Y}(H)$ for Algorithm \ref{alg:1}, is optimized in order to maximize the size of the estimate of the basin of attraction. This is different from  \citep{yin2021stability}, where regional sector bounds are a-priori fixed and then local stability properties are assessed.}

\begin{rmk_new}\label{rmk:model}
	\revSeven{The robustness of the stability properties of the equilibrium of the control system with respect to possible uncertainties is a key feature enabling the controller to be efficiently applied. In our case, robustness is guaranteed, besides by the inherent robustness of the stability of Lyapunov stable equilibria (see \citep[Chapter 7]{BookTeel}), by the fact that our conditions are not defined for the specific model \eqref{eq:sys1} solely, but for all nonlinearities $\sigma(\cdot)$ fulfilling the general sector condition specified in Assumption \ref{ass:1}.		
	Note that, while there may not be an exact correspondence between model \eqref{eq:sys1} and the real plant, the estimate basin of attraction gives a measure of the range of validity of the  regional stability properties. Even more so, through the output map \eqref{eq:sys1_out} it is possible to define a correspondence between such positively invariant set and the region of validity of our controller for the plant quantities; if necessary, the computation of $S$ can be done by maximizing, at the same time, the size of its projection onto the output space through a suitable choice of function $f_{\rm LMI}(\cdot)$.}
\end{rmk_new}

\begin{rmk_new}\label{rmk:obs}
	\revSeven{As discussed in Section \ref{sec:ProbStat}, the considered state feedback control scheme may necessitate the design of a suitable state observer, as typical for RNN models. The scheme would then be perturbed by a vanishing estimation error, which does not invalidate the established stability properties. Indeed, one may follow similar constructions to those of $\delta$ISS observers \citep{d2023data}, which can be easily designed for \eqref{eq:sys2} considering that $\sigma$, under Assumption \ref{ass:1}, is incrementally sector bounded.}
\end{rmk_new}
\section{Controller design}
\label{sec:controldesign}

As previously discussed, the conditions presented in Section \ref{sec:RegCond} are quite effective for analysing the regional stability of the origin for system \eqref{eq:sys2}, but they can also be an efficient tool for the design of stabilizing controllers. Hence, the objective addressed in this section is to show how to design a state-feedback control law for system \eqref{eq:sys2}, i.e., \eq{u_k=Kx_k}, such that: \textit{(i)} regional (or global) asymptotic stability of the origin is ensured, \textit{(ii)} an estimate of the basin of attraction is determined, and \textit{(iii)} desired closed-loop performances are enforced.
%
%
\revFive{To this end, }introduce \eq{F = A_{\circ} + B_{\sigma} C_\circ}, \eq{G=B_u + B_{\sigma}D_u}. 
\revFive{It is worth noting that, for achieving global stability of the closed loop, it is necessary that a $K$ exists such that both $A_\circ+B_uK$ and $F+GK$ are Schur with a common Lyapunov function (in view of Lemma \ref{lm:1}). On the other hand, for regional stability is enough that $K$ exists such that $F+GK$ is Schur (in view of Lemma \ref{lm:feasth1} and \ref{prop:2}).}\\
\revFive{Considering the mentioned objectives, }Theorem~\ref{th:1} or~\ref{th:2} can be exploited.  \revFive{Defining \eq{J=KS},} condition~\eqref{eq:th1_cond1} in Theorem~\ref{th:1} can be written as
\begin{align}\label{eq:th1_design}
	&\begin{bmatrix}
		S & -L-\Omega\T &-\Omega\T\Theta  &\Lambda\T\;\\
		-L-\Omega  & 2U &0 & UB\T\\
		-\Theta \Omega  & 0 &2R & RB\T\\
		\Lambda & BU & BR & S\\
	\end{bmatrix}\!\!\succ 0,
\end{align}
\ifred
where $\Omega =CS= C_\circ S + D_u J$ and 	$\Lambda= AS= F S + G J$.
\else
with the next linear functions of the decision variables, issued from the notation in \eqref{eq:sys2}:
\begin{align*}
	\Omega =CS= C_\circ S + D_u J, \quad
	\Lambda= AS= F S + G J\,.
\end{align*}
\fi
Similarly, condition~\eqref{eq:th2} in Theorem \ref{th:2} can be written as 
\begin{subequations}\label{eq:th2_design}
	\begin{align}
		&\begin{bmatrix}\label{eq:th2_cond1_design}
			S &-\Omega\T &\Lambda\T\\
			-\Omega & 2(H+I)U & UB\T\\
			\Lambda & BU & S
		\end{bmatrix}\succ 0,\\ \nonumber \\[-0.3cm]
		&\begin{bmatrix}\label{eq:th2_cond2_design}
			S &	\Omega_{i,:}\T\\
			\Omega_{i,:} & \bar{y}^2_i(h_i)
		\end{bmatrix}\succeq 0, \quad \forall i \in \{1,\hdots, \nu\}\,, 
	\end{align}
\end{subequations}
Besides stability, the state-feedback gain $K$ should be chosen to impose desired closed-loop performances.
Consider the linearization of \eqref{eq:sys2} around zero equilibrium \begin{align}\label{eq:syslin}
	x^+ = (F + GK)  x + d\,.
\end{align}
The term $d$ can be regarded as an exogenous disturbance or, recalling that \eq{d=Bq(y)}, it can account for neglected nonlinarities, whose effect can be attenuated by means of suitable minimization-based design approaches.
\revCommon{As later discussed, different control strategies can be combined with the conditions described in Section \ref{sec:RegCond} to enforce performance and stability properties to the closed-loop system via matrix inequalities. Here, 
the $\mathcal{H}_2$-based approach is leveraged, given its desirable noise rejection properties. In particular, $\mathcal{H}_2$ control  minimizes the effect of the exogenous disturbance $d$, representing the nonlinearity excess (see \eqref{eq:sys2}), on the performance output}
\begin{align}\label{eq:perf_h2}
	z = \widetilde{Q}x + \widetilde{R}u =  (\widetilde{Q} + \widetilde{R}\,K)  x\,,
\end{align}
where \eq{\widetilde{Q}}, \eq{ \widetilde{R}} are tunable output matrices. $K$ is designed to minimize the $\mathcal{H}_2$ norm of the transfer function from \revCommon{$d$} to $z$ in the region of linear behaviour, namely in the tail of the nonlinear closed-loop responses. The next proposition allows to cast the design as a quasi-convex optimization problem.

	\begin{prop_new}\label{prop:H2}
		The $\mathcal{H}_2$  control problem applied to \eqref{eq:syslin} can be recast as the solution to the following (quasi-convex) generalized eigenvalue problem
		\begin{subequations}\label{eq:H2_ext}
			\begin{align}
				& \qquad \qquad \underset{\substack{J\in\mathbb{R}^{m\times n}, \; \Gamma=\Gamma\T \in \mathbb{R}^{n\times n}\\ S \in \mathbb{S}^{n}_{\succ 0},\, \delta\in \mathbb{R}_{\geq 0},\, \eta\in \mathbb{R}_{\succ 0}}}{\min} \;\; \delta \nonumber
				\\
				&\text{subject}\text{ to} \nonumber \\
				&\begin{bmatrix}
					S& SF\T + J\T G\T &S \widetilde{Q}\T + J\T \widetilde{R}\T\\
					FS + GJ&S &0\\
					\widetilde{Q}S + \widetilde{R}J  &0& \eta \,I
				\end{bmatrix} \!\!	\succ\! 0, \label{eq:H2_matr1}\\
				&\begin{bmatrix}
					\Gamma &\eta I\\
					\eta I & S
				\end{bmatrix}\succeq 0,\label{eq:H2_matr2}\\
				&\;\,\mathrm{trace}(\Gamma) <  \eta\, \delta\,,\label{eq:H2_matr3}
			\end{align}	
		\end{subequations}
		where the state-feedback gain is selected as $K=JS^{-1}$.
	\end{prop_new}

\ifred
\else
\textbf{Proof of Proposition \ref{prop:H2}.} 
Considering \eqref{eq:syslin} and \eqref{eq:perf_h2}, when $F+GK$ is Schur stable, it is shown in \citep{van2020noisy} that the $\mathcal{H}_2$ norm of the transfer function from $w$ to $z$ is upper bounded by $\sqrt{\delta}$, if, and only if, there exists \eq{\widetilde{P}\in \mathbb{S}^{n}_{\succ 0}} such that
\begin{subequations}
	\begin{align}
		&\widetilde{P}\!\succ\! \!(F\!+\!GK)\!\!\T\! \widetilde{P} (F\!+\!GK) \!+\! (\widetilde{Q} \!+\! \widetilde{R}K)\!\T\, \!\!(\widetilde{Q}\! +\! \widetilde{R}K),\label{eq:vw_1}\\
		&\text{trace}(\widetilde{P}) < \delta\,. \label{eq:vw_2}
	\end{align}
\end{subequations}
\ifred
\RED{Multiplying \eqref{eq:vw_1} by $\eta^{-1}>0$ and by relying on the Schur complement,  \eqref{eq:vw_1} is equivalent to}
\else
Multiplying \eqref{eq:vw_1} by $\eta^{-1}>0$, we obtain the equivalent inequality
\begin{align}
&	\eta^{-1}\widetilde{P}-	\begin{bmatrix}
		(F\!+\!GK)\!\T  (\widetilde{Q}\! +\! \widetilde{R}K)\!\T
	\end{bmatrix}\!
	\begin{bmatrix}
		\eta^{-1}\widetilde{P} \quad 0\\ 0 \quad \eta^{-1}I
	\end{bmatrix}\!
	\begin{bmatrix}
		F\!+\!GK\\ \widetilde{Q}\! +\! \widetilde{R}K
	\end{bmatrix}\nonumber\\&\succ0\,,
\end{align}
which, by a Schur complement, is equivalent to
\fi
\begin{align}\label{eq:matrh2}
	\begin{bmatrix}
		\eta^{-1}\widetilde{P}\quad &(F\!+\!GK)\!\T \quad &(\widetilde{Q}\! +\! \widetilde{R}K)\!\T \;
		\\ F\!+\!GK \quad &\eta\widetilde{P}^{-1} \quad &0\\
		\widetilde{Q}\! +\! \widetilde{R}K \quad &0 \quad &\eta I
	\end{bmatrix}\succ0\,.
\end{align}
By defining $P=\eta^{-1}\widetilde{P}$ and $S=P^{-1}$, and by pre-multiplying and post-multiplying \eqref{eq:matrh2} by $M_4=\text{diag}(S,I,I)$, the inequality \eqref{eq:H2_matr1} is obtained, where $J=KS$.
To reformulate \eqref{eq:vw_2}, recall that $\widetilde{P}=\eta S^{-1}$. In view of this,  first note that, by \eqref{eq:H2_matr2}, $\Gamma$ satisfies $\Gamma \succeq \eta^2 S^{-1}$. Then,  \eqref{eq:vw_2}, i.e., trace($\eta^{-1}P)<\delta$, is equivalent to \eqref{eq:H2_matr3}. \hfill$\square$
\fi

Note that, if $\eta=1$, \eqref{eq:H2_ext} becomes an LMI-based problem similar to \citep{van2020noisy}. This would however restrict the degrees of freedom of the problem, limiting the size of the estimate of the basin of attraction $\mathcal{E}(S)$.
Due to Proposition \ref{prop:H2}, in order to guarantee desired performances, the local stability of the origin relying on Theorem~\ref{th:1}, and the maximization of the basin of attraction estimate, the following problem can be formulated
\begin{align}\label{eq:H2_th1}
			\begin{split}
				&\underset{\substack{S, \,\Gamma\in \mathbb{S}_{\succ 0}^{n},\, \,L\in \mathbb{R}^{\nu\times n}, U, R\, \in \mathbb{D}_{\succ 0}^{\nu}, \,\\ J\in\mathbb{R}^{m\times n}, \;\delta\in \mathbb{R}_{\geq0}, \eta \in \mathbb{R}_{\succ 0}} }{\min} \;\; \quad w\,\delta \; - \;\gamma
				\\
				& \text{subject} \text{ to }   \text{\eqref{eq:th1_cond2}, \eqref{eq:th1_design}, \eqref{eq:H2_ext}, } \text{and }  f_{\rm LMI}(S)\succeq \gamma I \,,
			\end{split}
		\end{align}
where $w\in \mathbb{R}_{\geq 0}$ is a design parameter, which can be tuned based on the prioritization of  the $\mathcal{H}_2$ norm minimization or the basin of attraction maximization, and $f_{\rm LMI}(S)$ a function selected by the designer as discussed in Section \ref{sec:condition1}.
		On the other hand, we can address the control design problem by resorting on Theorem \ref{th:2}. Given the nonlinearity \eqref{eq:th2_design}, a recursive procedure as the one sketched in Algorithm~\ref{alg:1} must be adopted. In particular, it is enough to solve, in step~5), the following alternative design problem for each $i$th iteration
		\begin{align}\label{eq:H2_th2}
			\begin{split}
				&\underset{\substack{J\in\mathbb{R}^{m\times n}, \, S,\Gamma \in \mathbb{S}^{n}_{\succ 0},\\ U \in \mathbb{D}^{\nu}_{\succ 0}, \delta\in \mathbb{R}_{>0},\, \eta \in \mathbb{R}_{>0}}} {\min} \;\;\;\; w\,\delta \; - \;\gamma
				\\
				&\text{subject} \text{ to } \qquad  \text{\eqref{eq:th2_design}, \eqref{eq:H2_ext}, }  \text{and }  f_{\rm LMI}(S)\succeq \gamma I \,,
			\end{split}
		\end{align}
with $w\in \mathbb{R}_{>0}$ being a design parameter, and  $f_{\rm LMI}(S)$ a suitably chosen function for the maximization of the estimate of the basin of attraction.
	The next main result is a straightforward consequence of Theorems \ref{th:1} and \ref{th:2}. 
	\begin{thm_new}\label{th:3}
		Any solution to \eqref{eq:H2_th1} or \eqref{eq:H2_th2} is such that the origin is exponentially stable for the closed-loop system \eqref{eq:sys2} with the selection $K=JS^{-1}$, with a basin of attraction containing the set $\mathcal{E}(S)$.
	\end{thm_new}
	\begin{rmk_new}
		Denoting by  $J^*,S^*$ the optimal solution either of \eqref{eq:H2_th1} or \eqref{eq:H2_th2}, it is worth noting that the resulting estimate of the basin of attraction, i.e., \eq{\mathcal{E}({S^*})=\{x:x\T{S^{*}}^{-1}x\leq 1 \}}, may be not the largest one, as this is not maximized in the $\mathcal{H}_2$  control problem. However, a larger basin of attraction can be computed by selecting the optimal state-feedback gain as $K=J^*{S^*}^{-1}$, and solving either \eqref{eq:gamma_th1} or Algorithm \ref{alg:1} following the analysis results in Section \ref{sec:RegCond}.
	\end{rmk_new}
	\begin{rmk_new}\label{rmk:analysis}
		Other control design strategies can be combined with the proposed regional stability conditions, leading to LMI control design problems, such as $\mathcal{H}_\infty$, using ideas from \citep{van2020noisy}, or \revSeven{Virtual Reference Feedback Tuning} (VRFT), using ideas from \citep{williamIFAC2023}.
	\end{rmk_new}

	\section{Numerical results}\label{sec:num}
\begin{figure}[t!]
		\centering
		\includegraphics[trim=50 10 25 10,clip=true,width=0.4\linewidth]{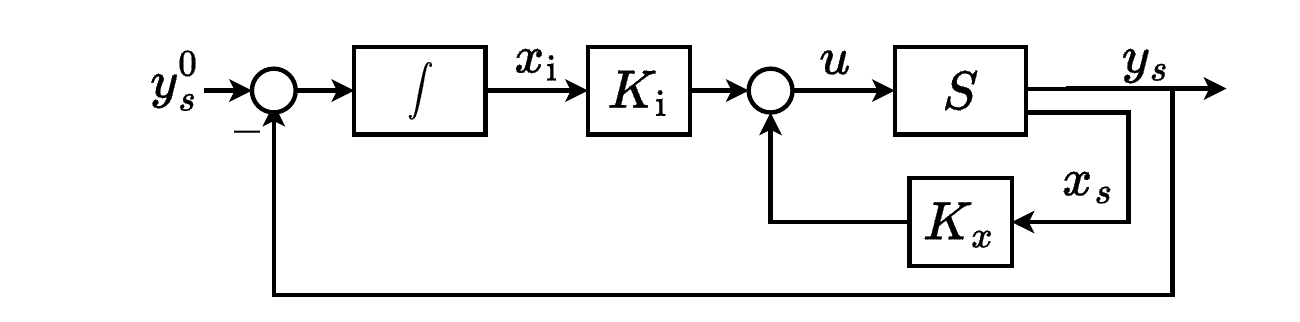}
		\caption{State-feedback control scheme with integral action.}
		\label{fig: SIMCS}
	\end{figure}

\begin{figure}[t!]
		\centering
		\subfloat[\empty]{\includegraphics[trim=5 10 5 10,clip=true, width=0.34\linewidth]{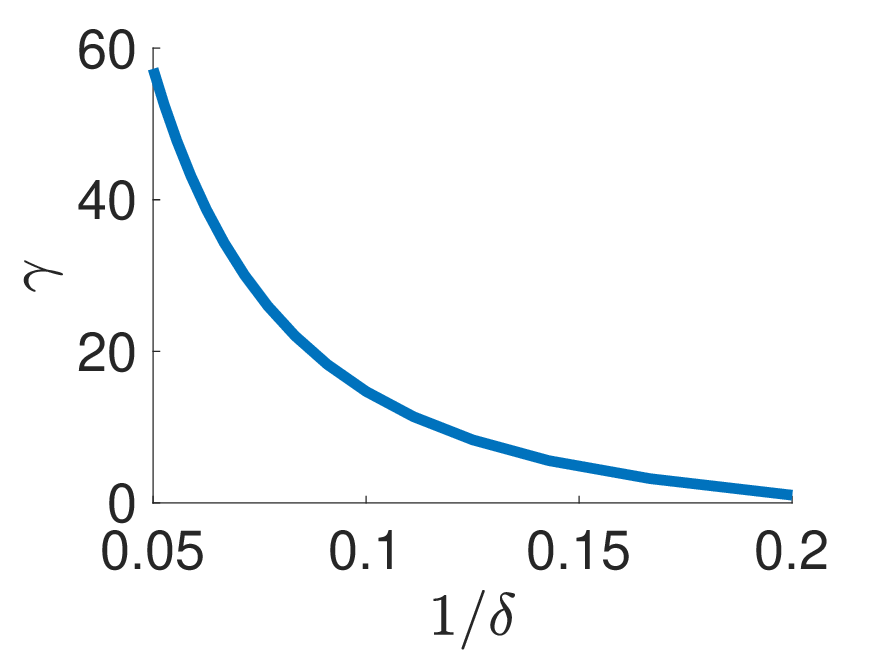}}\qquad \qquad
		\subfloat[\empty]{\includegraphics[trim=5 10 5 10,clip=true, width=0.34\linewidth]{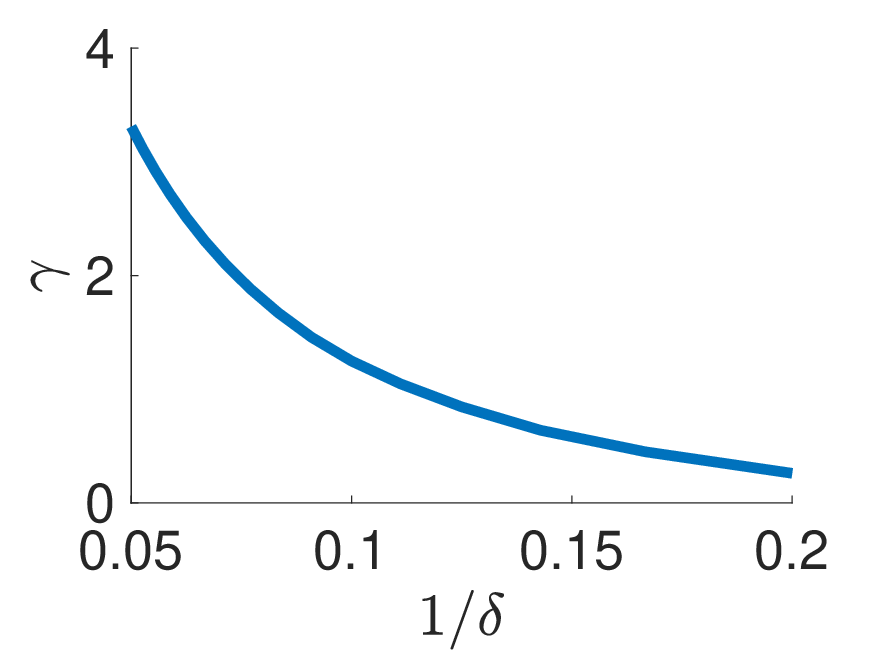}}
		\caption{Radius of the ellipsoid $\mathcal{E}(S)$, i.e., $\gamma$, for varying $\mathcal{H}_2$ norm $\bar{\delta} \in [5,20]$. Left plot: solution obtained from \eqref{eq:H2_th1}, right plot: solution obtained from \eqref{eq:H2_th2}.} 
		\label{fig:Pareto}
	\end{figure}

\begin{figure}[t!]
		\centering
		\subfloat[\empty]{\includegraphics[trim=30 10 5 30,clip=true, width=0.34\linewidth]{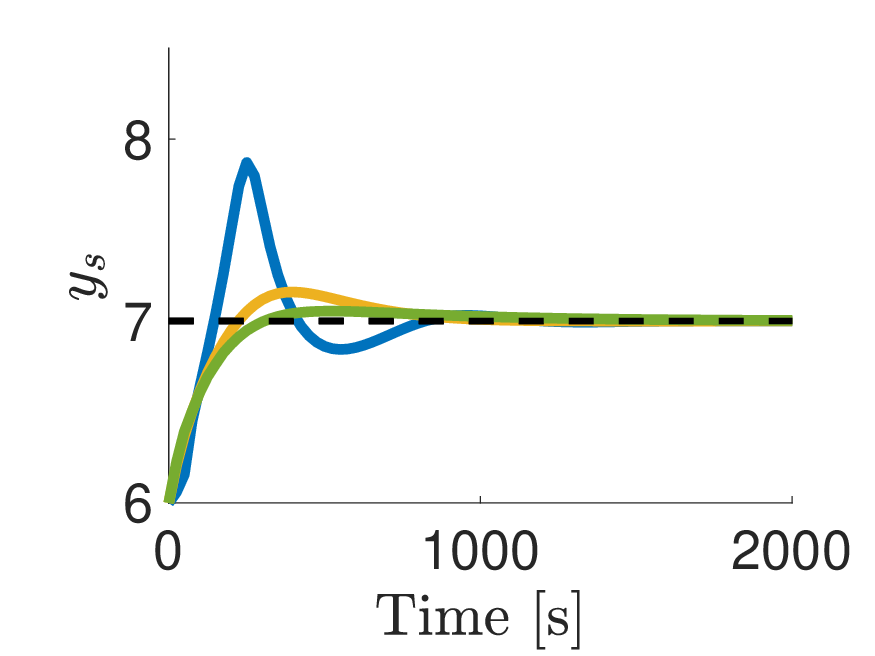}}\qquad \qquad
		\subfloat[\empty]{\includegraphics[trim=30 10 5 30,clip=true, width=0.34\linewidth]{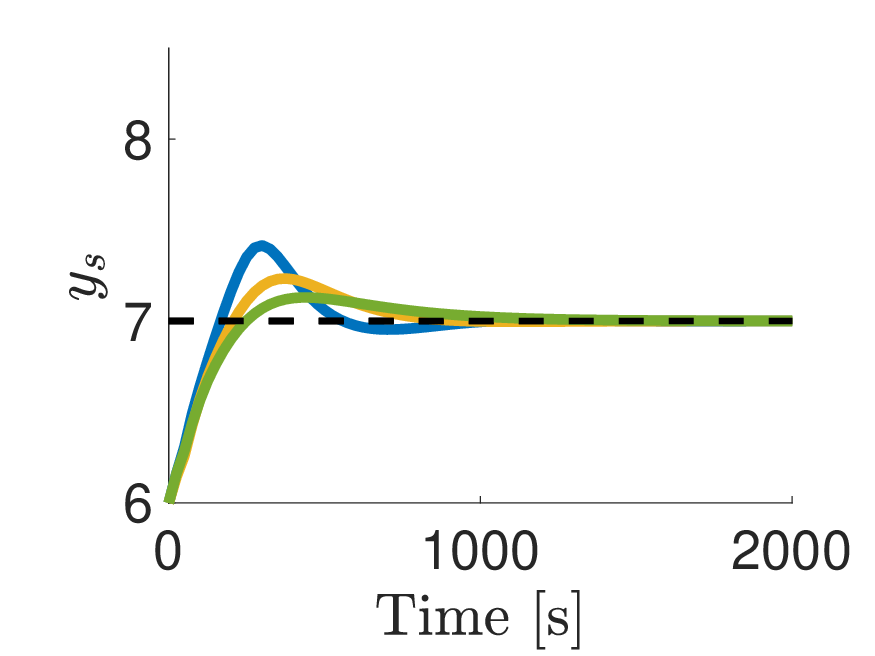}}
		\caption{Left plot: Output response with control law computed from \eqref{eq:H2_th1}, right plot: computed from \eqref{eq:H2_th2}  for $\bar{\delta}=5$ (blue), $\bar{\delta}=10$ (yellow), $\bar{\delta}=20$ (green).}
		\label{fig:Transients}
	\end{figure}

	The proposed approach is tested on a realistic benchmark taken from the literature, representing a {pH} neutralization process; for details see \citep{armenio2019model}. The {pH} process is a nonlinear SISO dynamical system where the input is the alkaline flowrate and the output is the {pH} concentration of a chemical solution. This system is simulated on laptop with an Intel i7-6500U processor, whereas the optimization problems are solved using Matlab with MOSEK as a solver.  \\
	First of all, the system model is exploited to generate input-output data through different experiments. In particular, 30000 samples of data are collected with a sampling time $T_s=25$ s, using a multilevel pseudo-random signal (MPRS) for the input (i.e., the alkaline flowrate) with amplitude in the range $[14.35, 16.35]$ mL/s.
	This dataset is exploited to derive a ESN model identifying the system dynamics, having structure
	\begin{align}\label{esnsim}
		\begin{cases}
			x_{s}^+ &\!\!\! = \sigma_{s}(W_{x}\,x_{s}+W_{u}\,u+W_{xy}\,y_{s}) \,,
			\\
			y_{s} &\!\!\!= W_{y}\,x_{s}\,,
		\end{cases}	
	\end{align}
	where $x_s\in \mathbb{R}^{n_s}$, $u\in \mathbb{R}$ and $y_s\in \mathbb{R}$ are the ESN state, input, and output vectors, respectively,  the function $\sigma_s:\mathbb{R}^{n_s} \mapsto \mathbb{R}^{n_s}$ has components  $\sigma_{{s},i}(\cdot)=\tanh(\cdot)$ for all $i=1,\dots,n_s$, whereas $W_{x}$, $W_{u}$,  $W_{xy}$, and  $W_{y}$ are the weighting matrices.
	The latter are tuned exploiting standard training procedures for ESNs, see \citep{armenio2019model} for more details, selecting $70\%$ of data as training set and $30\%$ as validation set.
The model order has been selected as $n_s=3$, \revSeven{which achieves a FIT equal to 73\%.} 
Note that the control design problem \eqref{eq:H2_th1}  displays numerical problems for ESN models with $n_s>3$. On the other hand, the control design problem  \eqref{eq:H2_th2} was always able to find a feasible solution for any of the tested number of ESN states with  $n_s \in \{3, 50\}$. \revSeven{Note that advanced techniques could be exploited for solving large-scale matrix problems, such as those based on matrix sparsity properties  \citep{zhang2018efficient} or on second order cone programming  \citep{ahmadi2017optimization}.} \smallskip\\
	%
	%
	%
	The objective of the control design procedure is to tune the gains $K_x$ and $K_\text{i}$ of the state-feedback and integrator which are, respectively, components of the control input $u$, as shown in Figure~\ref{fig: SIMCS}.	The integrator has model
	\ifred
	$	x_{\text{i}}^+ = \,x_{\text{i}} + e =  x_{\text{i}} + y^0_{s} - y_{s}$, 
	where $y_s^0$ is the output reference to be tracked and $e=y_s^0-y_s$ is the tracking error.
	Defining the closed-loop state vector as \mbox{$x=[x_s\T ,\,  x_{\text{i}}]\T$}  and the control input with a state-feedback control law, i.e., \mbox{$u = [K_x\; K_{\text{i}}]\,x = K x $}, the overall closed-loop system model can be written in the form 
	\else
	\begin{align}\label{eq:int}
		x_{\text{i}}^+ &= \,x_{\text{i}} + e =  x_{\text{i}} + y^0_{s} - y_{s}\,,
	\end{align}
	where $y_s^0$ is the output reference to be tracked and $e=y_s^0-y_s$ is the tracking error.
	Defining the closed-loop state vector as \mbox{$x=[x_s\T ,\,  x_{\text{i}}]\T$}  and the control input with a state-feedback control law, i.e., \mbox{$u = [K_x\; K_{\text{i}}]\,x = K x $}, the overall closed-loop system model \eqref{esnsim}-\eqref{eq:int} can be written in the form 
	\fi
	\begin{align}
		\label{eq:closedloop}
		\begin{cases}
			\begin{split}
				\!\!\!\!	x^+&= (F+G K)x+B\,q(y) + B_r\, y^0_\revSeven{s}\,,\\[0.2cm]
				y &= C x\,,
			\end{split}
		\end{cases}
	\end{align}
\ifred
where \mbox{$F\! =\! \begin{bmatrix}
			W_{x}+W_{xy}W_y & 0\\
			\revSeven{-W_{y}} & 1
		\end{bmatrix}$}\!,  \mbox{$G = [	W_{u}^\top \;	0]^{\!\top}\!\!$}, \mbox{$B = [-I \;0]^{\!\top}\!\!$}, \mbox{$B_r =[0 \;\;1]^{\!\top}\!\!$}, \mbox{$C = [W_x\!+\! W_{xy}W_y\! +\!W_u K_x \;\; W_u K_{\!\Scale[0.7]{\text{i}}} ]$},
\else
	where
	\begin{align*}
		&F = \begin{bmatrix}
			W_{x}+W_{xy}W_y & 0\\
			\revSeven{-W_{y}} & 1
		\end{bmatrix}, \quad G = \begin{bmatrix}
			W_{u}\\
			0 
		\end{bmatrix},\quad B = \begin{bmatrix}
			-I\\
			0
		\end{bmatrix},\\
		& B_r = \begin{bmatrix}
			0\\
			1
		\end{bmatrix}, \quad C =  \begin{bmatrix}
			W_x+ W_{xy}W_y +W_u K_x & W_u K_{\!\Scale[0.7]{\int}} 
		\end{bmatrix},
	\end{align*}
\fi
	and $q(y) = y - \sigma_s(y)$. It is apparent that \eqref{eq:closedloop} has the same structure as \eqref{eq:sys2}, apart from the presence of the reference input $y^0_s$, which has no influence on the stability and control design conditions described in the previous sections.
\ifred
Note that the ESN state $x_s$ in \eqref{esnsim} is not measurable in practice. Hence, a state observer is employed and designed as discussed in \citep{williamIFAC2023}, enjoying $\delta$ISS, and thus, vanishing estimation error (see also Remark \ref{rmk:obs}).
\else
	Note that the ESN state $x_s$ in \eqref{esnsim} is not measurable in practice. Hence, a state observer is employed, defined as
	\begin{align}	\label{eq:esnallobs}
		\begin{cases}
			\hat{x}_{s}^+ & \!\!\!\!= \sigma_{s}(W_{x}\hat{x}_{s}+\!W_{u}u+\!W_{xy}\hat{y}_{s} + \!\hat{L}(y_{s}-\hat{y}_{s})),\\
			\hat{y}_{s} &\!\!\!\!= W_{y}\,\hat{x}_{s}\,,
		\end{cases}
	\end{align}
	with $\hat{L}\in \mathbb{R}^{n}$. As discussed in \citep{williamIFAC2023}, $\hat{L}$ can be designed so that \eqref{eq:esnallobs} enjoys an \textit{incremental input-to-state stability} property, implying that $\hat{x}_s$ asymptotically converges to $x_s$.	
	\fi
	In the setting above, the control design solutions \eqref{eq:H2_th1} and \eqref{eq:H2_th2} are applied, so as to compute an optimal gain $K$ ensuring $\mathcal{H}_2$ performance and the stability of the origin for the nonlinear closed-loop system \eqref{eq:closedloop}. As for Theorem \ref{th:3}, the $\mathcal{H}_2$ performance output $z$  is selected as 
	\ifred
$		z_k = [ \sqrt{\tilde{q}_y} y^\top \;\sqrt{\tilde{q}_\text{i}} x_{\text{i}}^\top \sqrt{\tilde{r}_u} u^\top ]^\top.$	\else
	\begin{align}
		z_k = 
		 \begin{bmatrix}
			\tilde{q}_y^{\frac{1}{2}} \;y\\[0.2cm]
			\tilde{q}_{\text{i}}^{\frac{1}{2}} \,x_{\text{i}}\\[0.2cm]
			\tilde{r}_u^{\frac{1}{2}}\; u
		\end{bmatrix}\,,
	\end{align}
	where the weights $\tilde{q}_y$, $\tilde{q}_{\text{i}}$, $\tilde{r}_u>0$ are design parameters, with  formulation \eqref{eq:perf_h2}, this corresponds to selecting
	\begin{align}
		\widetilde{Q}= 
		\begin{bmatrix}
			\tilde{q}_y^{\frac{1}{2}}\, W_y & 0\\
			0 & \tilde{q}_{\text{i}}^{\frac{1}{2}}\\
			0 & 0
		\end{bmatrix}\,,\qquad  	\widetilde{R}= 
		\begin{bmatrix}
			0\\
			\tilde{r}_u^{\frac{1}{2}}
		\end{bmatrix}.
	\end{align}
\fi
	In particular, the $\mathcal{H}_2$ performance output is selected so as to minimize the output tracking error and the integrator state through the weights $\tilde{q}_y$ and $\tilde{q}_{\text{i}}$, as well as the control input effort through the weight $\tilde{r}_u$. 	
Thus, the proposed $\mathcal{H}_2$ control problem is solved by setting $\tilde{q}_y=\tilde{q}_{\text{i}}=0.1$, $\tilde{r}_u=0.05$, and $w=1$. 
Note that $\mathcal{H}_2$  control problem applied in combination with the global stability condition \eqref{eq:globalLMI} is not able to find a feasible solution. This is due to the presence of the system integrator and the bounded characteristic of the output in \eqref{esnsim}, as discussed in \citep{WilliamArxiv}. Thus, regional conditions are necessary, which can also define the maximum basin of attraction where the closed-loop stability property holds, by relying on the control problems \eqref{eq:H2_th1} and \eqref{eq:H2_th2}, which are solved by setting \eq{f_{\rm LMI}(S)=S}. As apparent from the cost functions used in~\eqref{eq:H2_th1} and~\eqref{eq:H2_th2}, the proposed design problems are multi-objective. In this section, to properly analyse the trade-off between the achieved performances, evaluated in terms of the $\mathcal{H}_2$ norm, and the achieved dimension of the basin of attraction, evaluated in terms of the radius of $\mathcal{E}(S)$, the control design problems \eqref{eq:H2_th1} and \eqref{eq:H2_th2} are solved by fixing the value of $\delta$, i.e., setting $\delta = \bar{\delta}$ with $\bar{\delta}\in [5,20]$.
	The achieved results are depicted in Figure \ref{fig:Pareto}, where the trade-off between the performances that can be requested to the control system and the achieved basin of attraction becomes apparent. In fact, the larger $1/\delta$, i.e., the smaller the $\mathcal{H}_2$ norm, the smaller the radius of the estimate $\mathcal{E}(S)$ of the basin of attraction.
	Also, notably, the control solution \eqref{eq:H2_th1} attains a larger estimate of the basin of attraction, with comparable $\mathcal{H}_2$ norm.
	Finally, the designed closed-loop controllers are tested on the real pH process model, considering a reference to be tracked at \mbox{$y_k^0=7$}; the results are reported in Figure \ref{fig:Transients}. The latter shows that (for the controllers obtained with \eqref{eq:H2_th1} and \eqref{eq:H2_th2}) the lower the $\mathcal{H}_2$ norm the smoother the output response, with reduced overshoots in the output transient. 
	\section{Conclusions}\label{sec:conclusion}
In this paper novel global and regional stability analysis conditions based on LMIs have been proposed for a general class of RNNs. 
	In particular, two alternative conditions are studied; the first one is obtained by defining an auxiliary function, while the second one is obtained by a sector narrowing technique.	We have shown how to leverage these conditions for state-feedback control design by formulating a suitable optimization problem enforcing $\mathcal{H}_2$ norm minimization.
	The theoretical properties of the methods presented here have been thoroughly characterized. 
	Finally, numerical simulations showed the advantages and limitations of the two methods. 
In particular, among the two regional stability conditions, the former one (based on the auxiliary function) \revOne{provides a larger estimate of the basin of attraction as compared to the latter one (based on sector narrowing) when both are feasible, but numerical tests show that the former condition may lead to numerical issues when high-dimensional systems are employed. On the other hand, the condition based on sector narrowing, despite returning a smaller estimate of the basin of attraction, manages to find a control design solution with stability guarantees also when large-order systems are considered.}
	This trade-off will be studied in future works.
	Future research will also include the extension of the obtained conditions to more complex control systems, possibly including nonlinear regulators belonging to more advanced RNN architectures (e.g., LSTM or GRU).
	
{	
\bibliographystyle{agsm}
\bibliography{Bibliografia.bib}
}
\end{document}